\documentstyle[amssymb,11pt,epsf]{article}

\begin{document}

\title{{\bf Measuring effective electroweak couplings in single top production at
the LHC}}
\author{D. Espriu\thanks{%
espriu@ecm.ub.es} \ and J. Manzano\thanks{%
manzano@ecm.ub.es} \\
\\
Departament d'Estructura i Constituents \\
de la Mat\`{e}ria and IFAE,\\
Universitat de Barcelona, \\
Diagonal, 647, E-08028 Barcelona}
\date{}
\maketitle

\begin{abstract}
We study the mechanism of single top production at the LHC in the framework
of an effective electroweak Lagrangian, analyzing the sensitivity of
different observables to the magnitude of the effective couplings that
parametrize new physics beyond the Standard Model. The observables relevant
to the distinction between left and right effective couplings involve in
practice the measurement of the spin of the top and this can be achieved
only indirectly by measuring the angular distribution of its decay products.
We show that the presence of effective right-handed couplings implies that
the top is not in a pure spin state. A unique spin basis is singled out
which allows one to connect top decay products angular distribution with the
polarized top differential cross section. We present a complete analytical
expression of the differential polarized cross section of the relevant
perturbative subprocess including general effective couplings. The mass of
the bottom quark, which actually turns out to be more relevant than naively
expected, is retained. Finally we analyze different aspects the total cross
section relevant to the measurement of new physics through the effective
couplings. The above analysis also applies to anti-top production in a
straightforward way.
\end{abstract}

\vfill
\vbox{
UB-ECM-PF 01/03\null\par
June 2001\null\par
}

\clearpage

\section{Introduction}

The standard model of electroweak and strong interactions has been, to this
day, tested to a remarkable degree of accuracy, particularly in what
concerns the neutral current sector. However it is clear that suffers from
several theoretical drawbacks (naturalness, triviality,...) making it
conceivable that it should be considered as an effective theory valid only
at low energies ($\lesssim 1$ TeV ). With the current limit on the Higgs
mass already placed at 113,5 GeV \cite{LEPC} and no clear evidence for the
existence of an elementary scalar (despite much controversy regarding the
results of the last days of LEP) it makes sense to envisage an alternative
to the minimal Standard Model described by an effective theory without any
physical light scalar fields. This in spite of the seemingly good agreement
between experiment and radiative corrections computed in the framework of
the minimal Standard Model (see \cite{chanowitz} however). This effective
theory should contain an infinite set of effective operators, of increasing
dimensionality, compatible with the electroweak and strong symmetries $%
SU(3)_{c}\times SU(2)_{L}\times U\left( 1\right) _{Y}$ and their
coefficients would parametrize physics beyond the Standard Model. In this
framework \cite{eff-lag} one can describe the low energy physics of theories
exhibiting the pattern of symmetry breaking $SU(2)_{L}\times U\left(
1\right) _{Y}\rightarrow U\left( 1\right) _{em}$ with full generality, in
the understanding, that this approach is useful as long as those particles
not explicitly included in the effective Lagrangian are much heavier than
the scale of energies at which the effective Lagrangian is to be used.

In this work we plan to investigate the new features that physics beyond the
standard model may introduce in the production of the top (or anti-top)
quarks through $W$-gluon fusion at the LHC. When describing the appropriate
effective vertex in the effective Lagrangian language, we will keep only the
leading non-universal (i.e. not appearing in the standard model) effective
operators in the low energy expansion. They correspond to the operators of
dimension four, which were first classified by Appelquist et al. \cite{appel}%
. These operators realize the $SU(2)_{L}\times U(1)_{Y}$ symmetry
non-linearly and are thus characteristic of strongly coupled theories and,
strictly speaking, are absent in the minimal Standard Model and
modifications thereof containing only light fields, such as supersymmetric
extensions, where all degrees of freedom are weakly interacting and need to
be included explicitly. When particularizing to interactions involving the $%
W,Z$ bosons, the operators present in the effective electroweak Lagrangian
induce effective vertices coupling the gauge boson to the matter fields (see
e.g. \cite{burgess}) 
\begin{equation}
-\frac{e}{4c_{W}s_{W}}\bar{f}\gamma ^{\mu }\left( \kappa _{L}^{NC}L+\kappa
_{R}^{NC}R\right) Z_{\mu }f-\frac{e}{s_{W}}\bar{f}\gamma ^{\mu }\left(
\kappa _{L}^{CC}L+\kappa _{R}^{CC}R\right) \frac{\tau ^{-}}{2}W_{\mu
}^{+}f+h.c.
\end{equation}
Other possible effects are not physically observable, as we shall see in a
moment. In practical terms, LHC will set bounds on these effective $W$
vertices, and therefore our results are also relevant in a broader
phenomenological context as a way to bound $\kappa _{L}$ and $\kappa _{R}$,
without any need to appeal to an underlying effective Lagrangian describing
a specific model of symmetry breaking. Of course one then looses the power
of an effective Lagrangian, namely a well defined set of counting rules and
the ability to relate different processes.

Another point to note is that, even in the minimal Standard Model, radiative
corrections induce modifications in the vertices. Assuming a smooth
dependence in the external momenta these form factors can be expanded in
powers of momenta. At the lowest order in the derivative expansion, the
effect of radiative corrections can be encoded in the effective vertices $%
\kappa _{L}$ and $\kappa _{R}$. Thus these effective vertices take well
defined, calculable values in the minimal Standard Model, and any deviation
from these values (which, incidentally, have not been yet fully determined
in the Standard Model yet) would indicate the presence of new physics in the
matter sector. The extent to what LHC can set direct bounds on the effective
vertices, in particular on those involving the third generation, is highly
relevant to constraint physics beyond the Standard Model in a direct way.
This paper is devoted to such an analysis in charged processes involving a
top quark at the LHC.

At the LHC energy (14 TeV) the dominant mechanism of top production, with a
cross section of 800 pb\cite{catani}, is gluon-gluon fusion. This mechanism
has nothing to do with the electroweak sector and thus is not the most
adequate for our purposes (although is the one producing most of the tops
and thus its consideration becomes necessary in order to study the top
couplings through their decay, which will not be our main interest here, and
also as a background to the process we shall be interested in).

Electroweak physics enters the game in single top production. (For a recent
review see e.g. \cite{Tait}.) At LHC energies the (by far) dominant
electroweak subprocess contributing to single top production is given by a
gluon ($g$) coming from one proton and a light quark or anti-quark coming
from the other (this process is also called $t$-channel production \cite
{tchannel,SSW}). This process is depicted in Figs. \ref{u+gt+b-d+tot} and 
\ref{d-gt+b-u-tot}, where light $u$-type quarks or $\bar{d}$-type antiquarks
are extracted from the proton, respectively. These quarks then radiate a $W$
whose effective couplings is the object of our interest. The cross total
section for this process at the LHC is estimated to be 250 pb\cite{SSW}, to
be compared to 50 pb for the associated production with a $W^{+}$ boson and
a $b$-quark extracted from the sea of the proton, and 10 pb corresponding to
quark-quark fusion ($s$-channel production). For comparison, at the Tevatron
(2 GeV) the cross section for $W$-gluon fusion is 2.5 pb, so the production
of tops through this particular subprocess is copious at the LHC. Monte
Carlo simulations including the analysis of the top decay products indicate
that this process can be analyzed in detail at the LHC and traditionally has
been regarded as the most important one for our purposes.

In a proton-proton collision a bottom-anti-top pair is also produced,
through the subprocesses given in Figs. \ref{d+gt-b+u+tot} and \ref
{u-gt-b+d-tot}. At any rate qualitative results are very similar to those
corresponding to top production, from where the cross sections can be easily
derived doing the appropriate changes.

In the context of effective theories, the contribution from operators of
dimension five to top production via longitudinal vector boson fusion was
estimated some time ago in \cite{LY}, although the study was by no means
complete. It should be mentioned that $t,\bar{t}$ pair production through
this mechanism is very much masked by the dominant mechanism of gluon-gluon
fusion, while single top production, through $WZ$ fusion, is expected to be
much suppressed compared to the mechanism presented in this paper, the
reason being that both vertices are electroweak in the process discussed in 
\cite{LY}, and that operators of dimension five are expected to be
suppressed, at least at moderate energies, by some large mass scale. The
contribution from dimension four operators as such has not, to our
knowledge, been considered before, although the potential for single top
production for measuring the CKM matrix element $K_{tb}$, has to some extent
been analyzed in the past (see e.g. \cite{SSW,mandp}).

\section{Effective couplings and observables}

Including family mixing and, possibly, $CP$ violation, the complete set of
dimension four effective operators which may contribute to the top effective
couplings and are relevant for the present discussion is \cite
{appel,Bagan,EspMan} 
\begin{eqnarray}
{\cal L}_{L}^{1} &=&i{\rm \bar{f}}M_{L}^{1}\gamma ^{\mu }U\left( D_{\mu
}U\right) ^{\dagger }L{\rm f}+h.c.,  \nonumber \\
{\cal L}_{L}^{2} &=&i{\rm \bar{f}}M_{L}^{2}\gamma ^{\mu }\left( D_{\mu
}U\right) \tau ^{3}U^{\dagger }L{\rm f}+h.c.  \nonumber \\
{\cal L}_{L}^{3} &=&i{\rm \bar{f}}M_{L}^{3}\gamma ^{\mu }U\tau
^{3}U^{\dagger }\left( D_{\mu }U\right) \tau ^{3}U^{\dagger }L{\rm f}+h.c., 
\nonumber \\
{\cal L}_{L}^{4} &=&i{\rm \bar{f}}M_{L}^{4}\gamma ^{\mu }U\tau
^{3}U^{\dagger }D_{\mu }^{L}L{\rm f}+h.c.,  \label{effL}
\end{eqnarray}

and 
\begin{eqnarray}
{\cal L}_{R}^{1} &=&i{\rm \bar{f}}M_{R}^{1}\gamma ^{\mu }U^{\dagger }\left(
D_{\mu }U\right) R{\rm f}+h.c.,  \nonumber \\
{\cal L}_{R}^{2} &=&i{\rm \bar{f}}M_{R}^{2}\gamma ^{\mu }\tau ^{3}U^{\dagger
}\left( D_{\mu }U\right) R{\rm f}+h.c.,  \nonumber \\
{\cal L}_{R}^{3} &=&i{\rm \bar{f}}M_{R}^{3}\gamma ^{\mu }\tau ^{3}U^{\dagger
}\left( D_{\mu }U\right) \tau ^{3}R{\rm f}+h.c.,  \label{effR}
\end{eqnarray}
where $L=\frac{1-\gamma ^{5}}{2},$ $R=\frac{1+\gamma ^{5}}{2}$ are the left
and right projectors, the matrices $M$ have family indices only, and the
above set is written the non-physical `weak' basis (that is, with matter
fields transforming as irreducible representations of the gauge group) \cite
{EspMan}. The unitary matrix $U$ contains the three Goldstone bosons
associated to the breakdown of the global symmetry $SU(2)_{L}\times
SU(2)_{R} $ down to $SU(2)_{V}$. The derivatives appearing in the effective
operators are given by 
\begin{eqnarray}
D_{\mu }U &=&\partial _{\mu }U+ig\frac{\tau }{2}\cdot W_{\mu }U-ig^{\prime }U%
\frac{\tau ^{3}}{2}B_{\mu },  \nonumber \\
D_{\mu }^{L}{\rm f} &=&\left( \partial _{\mu }+ig\frac{\tau }{2}\cdot W_{\mu
}+ig^{\prime }\left( Q-\frac{\tau ^{3}}{2}\right) B_{\mu }+ig_{s}\frac{%
\lambda }{2}\cdot G_{\mu }\right) {\rm f},  \nonumber \\
D_{\mu }^{R}{\rm f} &=&\left( \partial _{\mu }+ig^{\prime }QB_{\mu }+ig_{s}%
\frac{\lambda }{2}\cdot G_{\mu }\right) {\rm f},  \label{deriv}
\end{eqnarray}
where ${\rm f}$ is a weak doublet of matter fields with family indices also.
In addition, one has the following `universal' terms 
\begin{eqnarray}
{\cal L}_{SM} &=&i{\rm \bar{f}}\gamma ^{\mu }\left[ M^{L}D_{\mu
}^{L}L+\left( \tau ^{u}M_{u}^{R}+\tau ^{d}M_{d}^{R}\right) D_{\mu }^{R}R%
\right] {\rm f}  \nonumber \\
&&-{\rm \bar{f}}\left( U\left( \tau ^{u}\tilde{y}_{u}^{f}+\tau ^{d}\tilde{y}%
_{d}^{f}\right) R+\left( \tau ^{u}\tilde{y}_{u}^{f\dagger }+\tau ^{d}\tilde{y%
}_{d}^{f\dagger }\right) U^{\dagger }L\right) {\rm f},  \label{univer}
\end{eqnarray}
which are present in the Standard Model. In Eq.(\ref{univer}) we allow for
general couplings $M_{L}$, $M_{R}^{(u,d)}$; in the Standard Model these
couplings can be renormalized away via a change of basis, but in more
general theories they leave traces in other operators not present in the
Standard Model\cite{EspMan}.

In \cite{EspMan} it was also shown that when we diagonalize the mass matrix
present in Eq. (\ref{univer}) via a redefinition of the matter fields $%
\left( {\rm f}\rightarrow f\right) $ we change also the structure of
operators (\ref{effL},\ref{effR}). Taking that into account, the
contribution to the different gauge boson-fermion-fermion vertices is as
follows 
\begin{eqnarray}
{\cal L}_{bff} &=&-g_{s}\bar{f}\gamma ^{\mu }\left( a_{L}L+a_{R}R\right) 
\frac{{\bf \lambda }}{2}{\bf \cdot G}_{\mu }f,  \nonumber \\
&&-e\bar{f}\gamma ^{\mu }\left( b_{L}L+b_{R}R\right) A_{\mu }f,  \nonumber \\
&&-\frac{e}{2c_{W}s_{W}}\bar{f}\gamma ^{\mu }\left[ \left( c_{L}^{u}\tau
^{u}+c_{L}^{d}\tau ^{d}\right) L+\left( c_{R}^{u}\tau ^{u}+c_{R}^{d}\tau
^{d}\right) R\right] Z_{\mu }f  \nonumber \\
&&-\frac{e}{s_{W}}\bar{f}\gamma ^{\mu }\left[ \left( d_{L}L+d_{R}R\right) 
\frac{\tau ^{-}}{2}W_{\mu }^{+}+\left( d_{L}^{\dagger }L+d_{R}^{\dagger
}R\right) \frac{\tau ^{+}}{2}W_{\mu }^{-}\right] f,  \label{vert}
\end{eqnarray}
where $\tau ^{u}$ and $\tau ^{d}$ are the up and down projectors and $f$
represents the matter fields in the physical, diagonal basis. It can be
shown that once the all the renormalization (vertex, $CKM$ elements,
wave-function) countertems are taken into account \cite{EspMan} we obtain $%
a_{L,R}=1$, $b_{L,R}=Q$; i.e. we have no contribution from the effective
operators to the vertices of the gluon and photon. For the $Z$ couplings we
get instead 
\begin{eqnarray}
c_{L}^{u} &=&1-2Qs_{W}^{2}-N^{1}-N^{1\dagger }+N^{2\dagger
}+N^{2}+N^{3}+N^{3\dagger },  \nonumber \\
c_{L}^{d} &=&-1-2Qs_{W}^{2}+K^{\dagger }\left( N^{1}+N^{1\dagger
}+N^{2\dagger }+N^{2}-N^{3}-N^{3\dagger }\right) K,  \nonumber \\
c_{R}^{u} &=&-2s_{W}^{2}Q+\tilde{M}^{1}+\tilde{M}^{1\dagger }+\tilde{M}^{2}+%
\tilde{M}^{2\dagger }+\tilde{M}^{3}+\tilde{M}^{3\dagger },  \nonumber \\
c_{R}^{d} &=&-2s_{W}^{2}Q+\tilde{M}^{1}+\tilde{M}^{1\dagger }-\tilde{M}^{2}-%
\tilde{M}^{2\dagger }-\tilde{M}^{3}-\tilde{M}^{3\dagger },  \label{Zcoupl}
\end{eqnarray}
where $K$ is the $CKM$ matrix, and the matrices $N$'s and $\tilde{M}$'s are
redefined matrices according to the results of \cite{EspMan} (the exact
relation of these matrices to the $M$'s of Eqs. (\ref{effL},\ref{effR}) has
no relevance for the present discussion). Finally for the charged couplings
we have 
\begin{eqnarray}
d_{L} &=&K+\left( -N^{1}-N^{1\dagger }+N^{2}-N^{2\dagger }-N^{3}-N^{3\dagger
}+N^{4}-N^{4\dagger }\right) K,  \nonumber \\
d_{R} &=&\tilde{M}^{1}+\tilde{M}^{1\dagger }+\tilde{M}^{2}-\tilde{M}%
^{2\dagger }-\tilde{M}^{3}-\tilde{M}^{3\dagger }.  \label{vertices}
\end{eqnarray}

Since the set of operators (\ref{effL}-\ref{effR}) is the most general one
allowed by general requirements of gauge invariance, locality and
hermiticity; it is clear that radiative corrections, when expanded in powers
of $p^{2}$, can be incorporated into them. In fact, such an approach has
proven to be very fruitful in the past. Once everything is included we are
allowed to identify the couplings $d_{L,R}$ with $\kappa _{LR}^{CC}$. In
this paper we shall be concerned with the bounds that the LHC experiments
will be able to set on the couplings $\kappa _{LR}^{CC}$, more specifically
on the entries $tj$ of these matrices (those involving the top). In the rest
of the article we do not consider mixing and we consider non-tree level and
new physics contributions only on the $tb$ effective couplings, therefore in
the numerical simulations we have taken 
\begin{eqnarray*}
d_{L} &=&diag\ (K_{ud},K_{cs},g_{L}), \\
d_{R} &=&diag\ (0,0,g_{R}).
\end{eqnarray*}
When we speak along the paper of the results for the Standard Model at tree
level we mean $g_{L}=1$, and $g_{R}=0$. However, even though numerical
results are presented considering only the $tb$ entry ($g_{L}$ and $g_{R}$),
since flavour indices and masses are kept all along in the analytical
expressions (see Appendix), the appropriate changes to include other entries
are immediate.

The effective couplings of the neutral sector (\ref{Zcoupl}) can be
determined from the $Z\rightarrow f~\bar{f}$ vertex\footnote{%
A 3 $\sigma $ discrepancy with respect to the Standard Model results, mostly
due to the right-handed coupling, remains in the $Z$ couplings of the $b$
quark to this date.} \cite{Bagan}, but at present not much is known from the 
$tb$ effective coupling. This is perhaps best evidenced by the fact that the
current experimental results for the (left-handed) $K_{tb}$ matrix element
give \cite{PDG} 
\begin{equation}
\frac{|K_{tb}|^{2}}{|K_{td}|^{2}+|K_{ts}|^{2}+|K_{tb}|^{2}}=0.99\pm 0.29.
\end{equation}
In the Standard Model this matrix element is expected to be close to 1. It
should be emphasized that these are the `measured' or `effective' values of
the CKM matrix elements, and that they do not necessarily correspond, even
in the Standard Model, to the entries of a unitary matrix on account of the
presence of radiative corrections. These deviations with respect to unitary
are expected to be small ---at the few per cent level at most--- unless new
physics is present. At the Tevatron the left-handed couplings are expected
to be eventually measured with a 5\% accuracy \cite{TEVA}. The present work
is a contribution to such an analysis in the case of the LHC experiments.

As far as experimental bounds for the right handed effective couplings is
concerned, the more stringent ones come at present from the measurements on
the $b\rightarrow s\gamma $ decay at CLEO \cite{CLEO}. Due to a $m_{t}/m_{b}$
enhancement of the chirality flipping contribution, a particular combination
of mixing angles and $\kappa _{R}^{CC}$ can be found. The authors of \cite
{LPY} reach the conclusion that $|{\rm Re}(\kappa _{R}^{CC})|\leq 0.4\times
10^{-2}$. However, considering $\kappa _{R}^{CC}$ as a matrix in generation
space, this bound only constraints the $tb$ element. Other effective
couplings involving the top remain virtually unrestricted from the data. The
previous bound on the right-handed coupling is a very stringent one. It is
pretty obvious that the LHC will not be able to compete with such a bound.
Yet, the measurement will be a direct one, not through loop corrections.
Equally important is that it will yield information on the $td$ and $ts$
elements too, by just replacing the $\bar{b}$ quark in Figs. \ref
{u+gt+b-d+tot},\ref{d-gt+b-u-tot} by a $\bar{d}$ or a $\bar{s}$ respectively.

Now we shall proceed to analyze the bounds that single top production at the
LHC can set on the effective couplings. This combined with the data from $Z$
physics will allow an estimation of the six effective couplings (\ref{Zcoupl}%
-\ref{vertices}) in the matter sector of the effective electroweak
Lagrangian. We will, in the present work limit ourselves to the
consideration of the cross-sections for production of polarized top quarks.
We shall not consider at this stage the potential of measuring top decays
angular distributions in order to establish relevant bounds on the effective
electroweak couplings. This issue merits a more detailed analysis, including
the possibility of detecting $CP$ violation \cite{DambESp}.

\section{The cross section}

\label{cross}In order to calculate the cross section $\sigma $ of the
process $pp\rightarrow t\bar{b}$ we have used the CTEQ4 set of structure
functions \cite{CTEQ4} to determine the probability of extracting a parton
with a given fraction of momenta from the proton. Hence we write
schematically 
\begin{equation}
\sigma =\sum_{q}\int_{0}^{1}\int_{0}^{1}f_{g}(y)f_{q}(x)\hat{\sigma}%
(xP_{1},yP_{2})dxdy,  \label{pdfs}
\end{equation}
where $f_{q}$ denote the parton distribution function (PDF) corresponding to
the partonic quarks and antiquarks and $f_{g}$ indicate the PDF
corresponding to the gluon. In Eq.(\ref{pdfs}) we have set the light quark
and gluon momenta to $xP_{1}$ and $yP_{1}$, respectively. ($P_{1}$ and $%
P_{2} $ are the four-momenta of the two colliding protons.) The
approximation thus involves neglecting the transverse momenta of the
incoming partons; the transverse fluctuations are integrated over by doing
the appropriate integrals over $k_{T}$. We have then proceeded as follows.
We have multiplied the parton distribution function of a gluon of a given
momenta from the first proton by the sum of parton distribution functions
for obtaining a $u$ type quark from the second proton. This result is then
multiplied by the cross sections of the subprocesses of Fig.\ref
{u+gt+b-d+tot}. We perform also the analogous process with the $\bar{d}$
type anti-quarks of Fig.\ref{d-gt+b-u-tot}. At the end, these two partial
results are add up to obtain the total $pp\rightarrow t\bar{b}$ cross
section.

Typically the top quark decays weakly well before strong interactions become
relevant, we can in principle measure its polarization state with virtually
no contamination of strong interactions (see e.g. \cite{parke} for
discussions this point and section 6). For this reason we have considered
polarized cross sections and provide general formulas for the production of
polarized tops or anti-tops. To this end one needs to introduce the spin
projector 
\[
\left( \frac{1+\gamma _{5}\not{n}}{2}\right) , 
\]
with 
\begin{eqnarray*}
n^{\mu } &=&\frac{1}{\sqrt{\left( p_{1}^{0}\right) ^{2}-\left( \vec{p}%
_{1}\cdot \hat{n}\right) ^{2}}}\left( \vec{p}_{1}\cdot \hat{n},p_{1}^{0}\hat{%
n}\right) , \\
\hat{n}^{2} &=&1,\qquad n^{2}=-1,
\end{eqnarray*}
as the polarization projector for a particle or anti-particle of momentum $%
p_{1}$ with spin in the $\hat{n}$ direction. The calculation of the
subprocesses cross sections have been performed for tops and anti-tops
polarized in an arbitrary direction $\hat{n}$. Later we have analyzed
numerically different spin frames defined as follows

\begin{itemize}
\item  Lab helicity frame: the polarization vector is taken in the direction
of the three momentum of the top or anti-top (right helicity) or in the
opposite direction (left helicity).

\item  Lab spectator frame: the polarization vector is taken in the
direction of the three momentum of the spectator quark jet or in the
opposite direction. The spectator quark is the $d$-type quark in Fig.\ref
{u+gt+b-d+tot} or the $\bar{u}$-type quark in Fig.\ref{d-gt+b-u-tot}.

\item  Rest spectator frame: like in the Lab spectator frame we choose the
spectator jet to define the polarization of the top or anti-top. Here,
however, we define $\hat{n}$ as $\pm $ the direction of the three momentum
of the spectator quark in the top or anti-top rest frame (given by a pure
boost transformation $\Lambda $ of the lab frame). Then we have $%
n_{r}=\left( 0,\hat{n}\right) $ in that frame and $n=\Lambda ^{-1}n_{r}$
back to the lab frame.
\end{itemize}

The calculation of the subprocess polarized cross-section we present is
completely analytical from beginning to end and the results are given in the
Appendix. Both the kinematics and the polarization vector of the top (or
anti-top) are completely general. Since the calculation is of a certain
complexity a number of checks have been done to ensure that no mistakes have
been made. The integrated cross section agrees well with the results in \cite
{SSW} when the same cuts, scale, etc. are used. The mass of the top is
obviously kept, but so is the bottom mass. The latter in fact turns out to
be more relevant than expected as we shall see in a moment. As we have
already discussed, the production of flavours other than ${\bar{b}}$ in
association with the top can be easily derived from our results.

In single top production a distinction is often made between $2\rightarrow 2$
and $2\rightarrow 3$ processes. The latter corresponds, in fact, to the
processes we have been discussing, the ones represented in Fig.\ref
{u+gt+b-d+tot}, in which a gluon from the sea splits into a $b$ $\bar{b}$
pair. In the $2\rightarrow 2$ process the $b$ quark is assumed to be
extracted from the sea of the proton, and both $b$ and $\bar{b}$ are
collinear. Of course since the proton has no net $b$ content, a $\bar{b}$
quark must be present somewhere in the final state and the distinction
between the two processes is purely kinematical. As is well known, when
calculating the total cross section for single top production a logarithmic
mass singularity \cite{SSW} appears in the total cross section due to the
collinear regime where the $b$ quark (and the $\bar{b}$) quark have $%
k_{T}\rightarrow 0$. This kinematic singularity is actually regulated by the
mass of the bottom; it appears to all orders in perturbation theory and a
proper treatment of this singularity requires the use of the
Altarelli-Parisi equations and its resumation into a $b$ parton distribution
function. It is thus indistinguishable of the so-called $2\rightarrow 2$
process. Clearly an appropriate cut in $k_{T}$ should allow us to retain the
perturbative regime of the $2\rightarrow 3$ process, while suppressing the $%
2\rightarrow 2$ one.

Two approaches can be used at this point. One ---advocated by Willenbrock
and coworkers \cite{SSW} is to focus on the low $p_{T}$ regime, to minimize
the contribution of the $t,\bar{t}$ background. One actually is interested
in processes where one does not see the $\bar{b}$ (resp. $b$) quark which is
produced in association with the $t$ quark (resp. $\bar{t}$), and therefore
sets an upper cut on the $p_{T}$ of the $\bar{b}$. For this one has to take
into account the $2\rightarrow 2$ process and, in particular, one must pay
attention not to double count the low $p_{T}$ region (for the $\bar{b}$ (or $%
b$) quark) of the $2\rightarrow 3$ process, which is already included via a $%
b$ PDF. The reason given in \cite{SSW} is to separate this single top
production process from the dominant $gg\rightarrow \bar{t}t$, which
eventually also produces a $b$ and a $\bar{b}$. We view this strategy as
somewhat risky since, first of all, the separation between the $2\rightarrow
3$ and $2\rightarrow 2$ is not a clear cut one and, on top of that, it is a
region where the cross section is rapidly varying so the results do depend
on the way the separation is done. Furthermore it relies on assuming that we
have a good knowledge of the $b$ PDF at that scale; one will be forced to
assume that it behaves largely like the $s$ PDF at the end of the day. On
top of that, that strategy does not completely avoid the background
originated in $\bar{t}t$ production. That is, when in the decaying $\bar{t}%
\rightarrow W^{-}\bar{b}\rightarrow \bar{u}d\bar{b}$ the $\bar{b}$ is missed
along with the $\bar{u}$-type anti-quark in which case the $d$-type quark is
taken as the spectator or when the $\bar{b}$ is missed along with the $d$%
-type quark in which case the $\bar{u}$-type anti-quark is taken as the
spectator.

On the other hand, measuring the $\bar{b}$ (or $b$ for anti-top production)
momenta will allow a better kinematic reconstruction of the individual
processes. This should allow for a separation from the dominant mechanism of
top production through gluon fusion. Setting a sufficiently high cut for the
jet energy and a good jet separation should be sufficient to avoid
contamination from $t,\bar{t}$ when one hadronic jet is missed. Finally, the
spin structure of the top is completely different in both cases due to the
chiral couplings in electroweak production. Therefore, according to this
philosophy ---stay perturbative---, we have implemented a lower cut of 30
GeV in the transversal momentum of the $\bar{b}$ (resp. $b$) in top (resp.
anti-top) production.

\section{A first look at the results}

We shall now present the results of our analysis. To calculate the total
event production corresponding to different observables we have used the
integrating montecarlo program VEGAS \cite{vegas}. We present results after
one year (defined as 10$^{7}$ seg.) run at full luminosity in one detector
(100 ${\rm fb}^{-1}$ at LHC).

The total contribution to the electroweak vertices $g_{L}$, $g_{R} $ has two
sources: the effective operators parametrizing new physics, and the
contribution from the universal radiative corrections. In the standard
model, neglecting mixing, for example, we have a tree level contribution to
the $\bar{t}W_{\mu }^{+}b$ vertex given by $-\frac{i}{\sqrt{2}}\gamma _{\mu
}gK_{tb}L$. Radiative corrections (universal and $M_{H}$ dependent) modify $%
g_{L}$ and generate a non zero $g_{R}$. These radiative corrections depend
weakly on the energy of the process and thus in a first approximation we can
take them as constant. Our purpose is to estimate the dependence of
different LHC observables on these total effective couplings and how the
experimental results can be used to set bounds on them. Assuming that the
radiative corrections are known, this implies in turn a bound on the
coefficients of the effective electroweak Lagrangian.

Let us start by discussing the experimental cuts. Due to geometrical
detector constraints we cut off very low angles for the outgoing particles.
The top, anti-bottom, and spectator quark have to come out with an angle of
less than 10 degrees and more than 170 degrees. These angular cuts
correspond to a cut in pseudorapidity $\left| \eta \right| <2.44$. In order
to be able to detect the three jets corresponding to the outgoing particles
we implements isolation cuts of 20 degrees between each other.

As already discussed we use a lower cut of 30 GeV in the $\bar{b}$ jet. This
reduces the cross section to less than one third of its total value, since
typically the $\bar{b}$ quark comes out in the same direction as the
incoming muon and a large fraction of them do not pass the cut. Similarly, $%
p_{T}> 20$ GeV cuts are set for the top and spectator quark jets. These cuts
guarantee the validity of perturbation theory and will serve to separate
from the overwhelming background of low $p_{T}$ physics. These values come
as a compromise to preserve a good signal, while suppressing unwanted
contributions. They are similar, but not identical to the ones used in \cite
{SSW} and \cite{mandp}. To summarize 
\begin{eqnarray}
{\rm detector~geometry~cuts} &:&10^{o}\leq \theta _{i}\leq 170^{o},\quad i=t,%
\bar{b},q_{s},  \nonumber \\
{\rm isolation~cuts} &:&20^{o}\leq \theta _{ij},\quad i,j=t,\bar{b},q_{s}, 
\nonumber  \label{cuts} \\
{\rm theoretical~cuts} &:&20~{\rm GeV}\leq p_{1}^{T},\quad 20~{\rm GeV}\leq
q_{2}^{T},\quad 30~{\rm GeV}\leq p_{2}^{T},
\end{eqnarray}
where $\theta _{t}$, $\theta _{\bar{b}},$ $\theta _{q_{s}}$ are the polar
angles with respect to the beam line of the top, anti-bottom and spectator
quark respectively; $\theta _{t\bar{b}}$, $\theta _{tq_{s}},$ $\theta _{\bar{%
b}q_{s}}$ are the angles between top and anti-bottom, top and spectator, and
anti-bottom and spectator, respectively. The momenta conventions are given
in Figs.\ref{u+gt+b-d+tot}-\ref{u-gt-b+d-tot}.

Numerically, the dominant contribution to the process comes from the diagram
where a $b$ quark is exchanged in the $t$ channel, but a large amount of
cancellation takes place with the crossed interference term with the diagram
with a top quark in the $t$ channel. The smallest contribution (but
obviously non-negligible) corresponds to this last diagram. It is then easy
to see, given the relative smallness of the $b$ mass, why the process is so
much forward.

Undoubtedly the largest theoretical uncertainty in the whole calculation is
the choice of a scale for $\alpha _{s}$ and the PDF's. We perform a leading
order calculation in QCD and the scale dependence is large. We have made two
different choices. We present some results with the scale $p_{T}^{cut}$ used
in $\alpha _{s}$ and the gluon PDF, while the virtuality of the $W$ boson is
used as scale for the PDF of the light quarks in the proton. When we use
these scales and compute, for instance, the total cross section above a cut
of $p_{T}=20$ GeV in the $\bar{b}$ momentum, we get an excellent agreement
with the calculations in \cite{SSW}. Most of our results are however
presented with a common scale $\mu ^{2}=\hat{s}$, $\hat{s}$ being the
center-of-mass energy squared of the $qg$ subprocess. The total cross
section above the cut is then roughly speaking two thirds of the previous
one, but no substantial change in the distributions takes place. It remains
to be seen which one is the correct choice.

From our Monte Carlo simulation for single top production at the LHC after 1
year of full luminosity and with the cuts given above we obtain the total
number of events. This number depends on the value of the effective
couplings and on the top polarization vector $n$ given in the frames defined
in section \ref{cross} . If we call $N\left( g_{L},g_{R},\hat{n}%
,(frame)\right) $ to this quantity, we obtain the following results 
\begin{eqnarray}
N\left( g_{L},g_{R},\hat{n}=\pm \frac{\vec{p}_{1}}{\left| \vec{p}_{1}\right| 
},(lab)\right) &=&g_{L}^{2}\times \left( 3.73\mp 1.31\right) \times
10^{5}+g_{R}^{2}\times \left( 3.54\pm .97\right) \times 10^{5}  \nonumber \\
&&+g_{L}g_{R}\times \left( -.237\mp .0283\right) \times 10^{5},  \nonumber \\
N\left( g_{L},g_{R},\hat{n}=\pm \frac{\vec{q}_{2}}{\left| \vec{q}_{2}\right| 
},(lab)\right) &=&g_{L}^{2}\times \left( 3.73\pm 2.22\right) \times
10^{5}+g_{R}^{2}\times \left( 3.54\mp 2.12\right) \times 10^{5}  \nonumber \\
&&+g_{L}g_{R}\times \left( -.237\mp 0.001\right) \times 10^{5},  \nonumber \\
N\left( g_{L},g_{R},\hat{n}=\pm \frac{\vec{q}_{2}}{\left| \vec{q}_{2}\right| 
},(rest)\right) &=&g_{L}^{2}\times \left( 3.73\pm 2.49\right) \times
10^{5}+g_{R}^{2}\times \left( 3.54\mp 2.15\right) \times 10^{5}  \nonumber \\
&&+g_{L}g_{R}\times \left( -.237\mp .0180\right) \times 10^{5},
\label{results}
\end{eqnarray}
where we have omitted the $\sqrt{N}$ statistical errors and we have
neglected possible $CP$ phases ($g_{L}$ and $g_{R}$ real). One can observe
from the simulations that the production of negative helicity (left) tops
represents the 69\% of the total single top production, this predominance of
left tops in the tree level electroweak approximation is expected due to the
suppression at high energies of right-handed tops because of the zero right
coupling in the charged current sector. In fact the production of
right-handed tops would be zero were it not for the chirality flip, due to
the top mass, in the $t$-channel. Of course the name `left' and `right' are
a bit misleading; we really mean negative and positive helicity states.
Chirality states cannot be used, because the production is peaked in the 200
to 400 GeV region for the energy of the top and the mass cannot be neglected.

\begin{figure}[tbp]
\epsfysize=10.cm 
\centerline{\epsffile{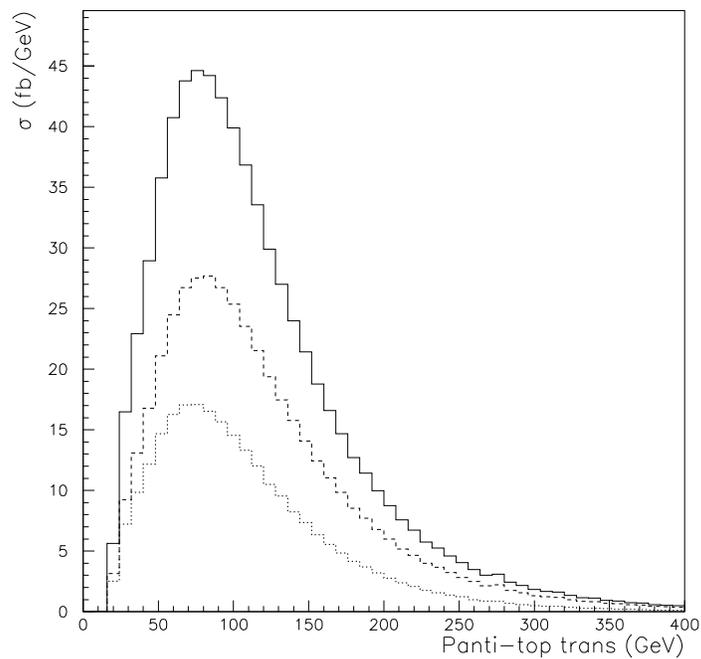}}
\caption{Anti-top transversal momentum distribution corresponding to
polarized single anti-top production at the LHC. The solid line corresponds
to unpolarized anti-top production and the dashed (dotted) line corresponds
to negative (positive) helicity anti-top production. Same conventions and
scale as in the previous figures. }
\label{patoptrans_lab_lr_gl=1_gr=0}
\end{figure}

\begin{figure}[tbp]
\epsfysize=10.cm 
\centerline{\epsffile{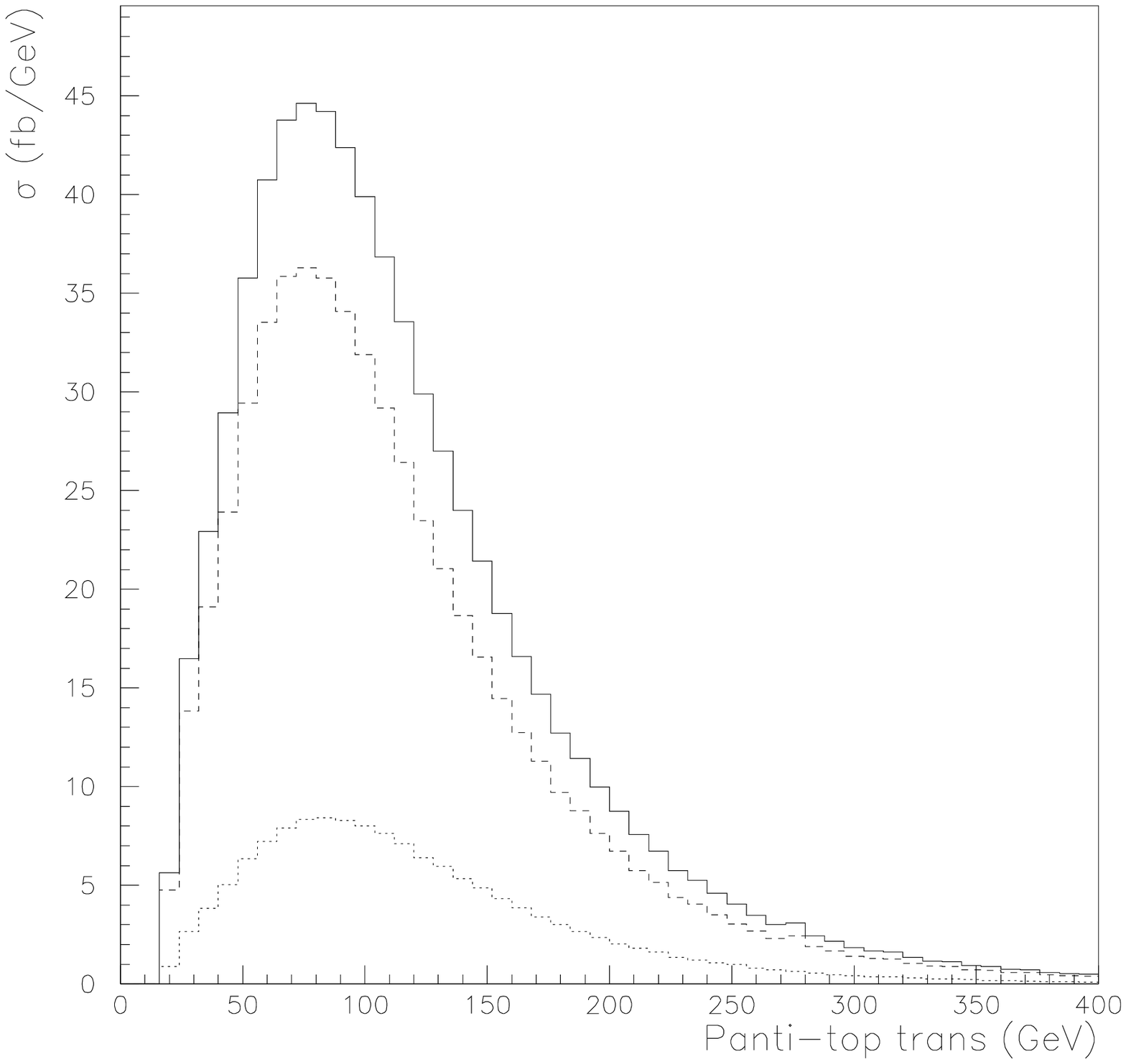}}
\caption{Anti-top transversal momentum distribution corresponding to
polarized single anti-top production at the LHC. The solid line corresponds
to unpolarized anti-top production and the dashed (dotted) line corresponds
to anti-tops polarized in the spectator jet negative (positive) direction in
the top rest frame. The subprocesses contributing to these histograms have
been calculated at tree level in the electroweak theory, using the same cuts
and conventions as in the previous figures.}
\label{patoptrans_rep_+-q2_gl=1_gr=0}
\end{figure}

We have also calculated single anti-top production. In Fig.\ref
{patoptrans_lab_lr_gl=1_gr=0} we show the histograms corresponding to the
production of $\bar{t}$ with the two possible helicities in the LAB frame.
All the histograms correspond to the tree level electroweak approximation
and clearly show that single anti-top production is approximately the 75\%
of single top production. The same pattern is observed in Fig. \ref
{patoptrans_rep_+-q2_gl=1_gr=0}. This suppression is generated by the parton
distribution functions corresponding to negatively charged quarks that are
smaller than the ones corresponding to positively charged quarks. Because of
that the conclusions for anti-top production are practically the same as the
ones for top production taking into account this suppression and that,
because of the transformations (\ref{change}) (see Appendix), passing from
top to anti-top is equivalent to changing the spin direction.

\begin{figure}[tbp]
\epsfysize=10.cm 
\centerline{\epsffile{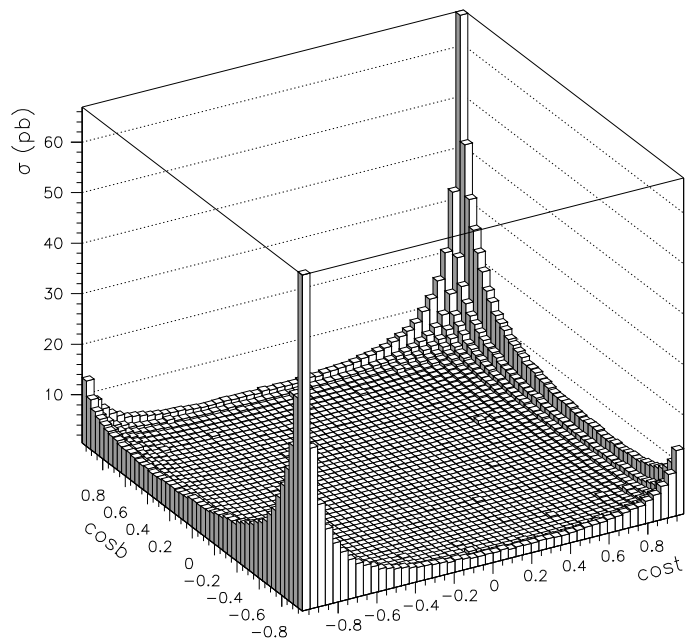}}
\caption{Distribution of the cosines of the polar angles of the top and
anti-bottom with respect to the beam line. The plot corresponds to
unpolarized single top production at the LHC. The calculation was performed
at the tree level in Standard Model with $\protect\mu ^{2}=\hat{s}=\left(
q_{1}+q_{2}\right) ^{2}$.}
\label{costcosb_lab_unpol_gl=1_gr=0}
\end{figure}

\begin{figure}[tbp]
\epsfysize=10.cm 
\centerline{\epsffile{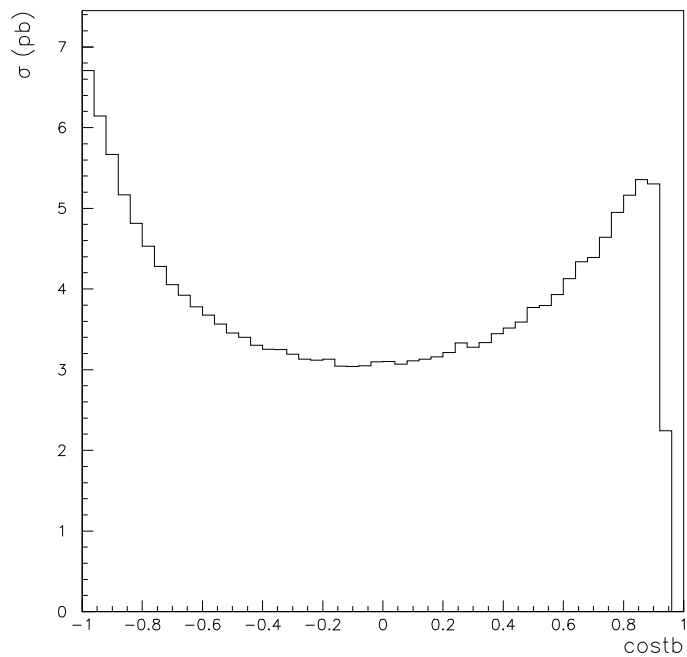}}
\caption{Distribution of the cosine of the angle between top and anti-bottom
corresponding to unpolarized single top production at the LHC. The
calculation was performed at the tree level in Standard Model with $\protect%
\mu ^{2}=\hat{s}=\left( q_{1}+q_{2}\right) ^{2}$. The abrupt fall near 1 is
due to the 20 degrees isolation cut.}
\label{costopbot_lab_unpol_gl=1_gr=0}
\end{figure}

In Fig. \ref{costcosb_lab_unpol_gl=1_gr=0} we plot the cross section
distribution of the polar angles of the top and anti-bottom with respect to
the beam line for unpolarized single top production at the LHC. In Fig. \ref
{costopbot_lab_unpol_gl=1_gr=0} we plot the distribution of the cosine of
the angle between the top and the anti-bottom for unpolarized single top
production at the LCH. Everything is calculated in the (tree-level) Standard
Model in the LAB frame. In both figures the above cuts are implemented, in
particular the isolation cut of 20 degrees in the angle between the top and
the anti-bottom is clearly visible in Fig. \ref
{costopbot_lab_unpol_gl=1_gr=0}. From inspection of these figures two facts
emerge: a) the distribution is strongly peaked in the beam direction as
expected. b) Even with the presence of the isolation cut, near the beam axis
configurations with top and anti-bottom almost parallel are favoured with
respect to back-to-back configurations. Therefore this is an indication that
almost back-to-back configurations are distributed more uniformly in space
than parallel configurations favoring the beam line direction.

\begin{figure}[tbp]
\epsfysize=10.cm 
\centerline{\epsffile{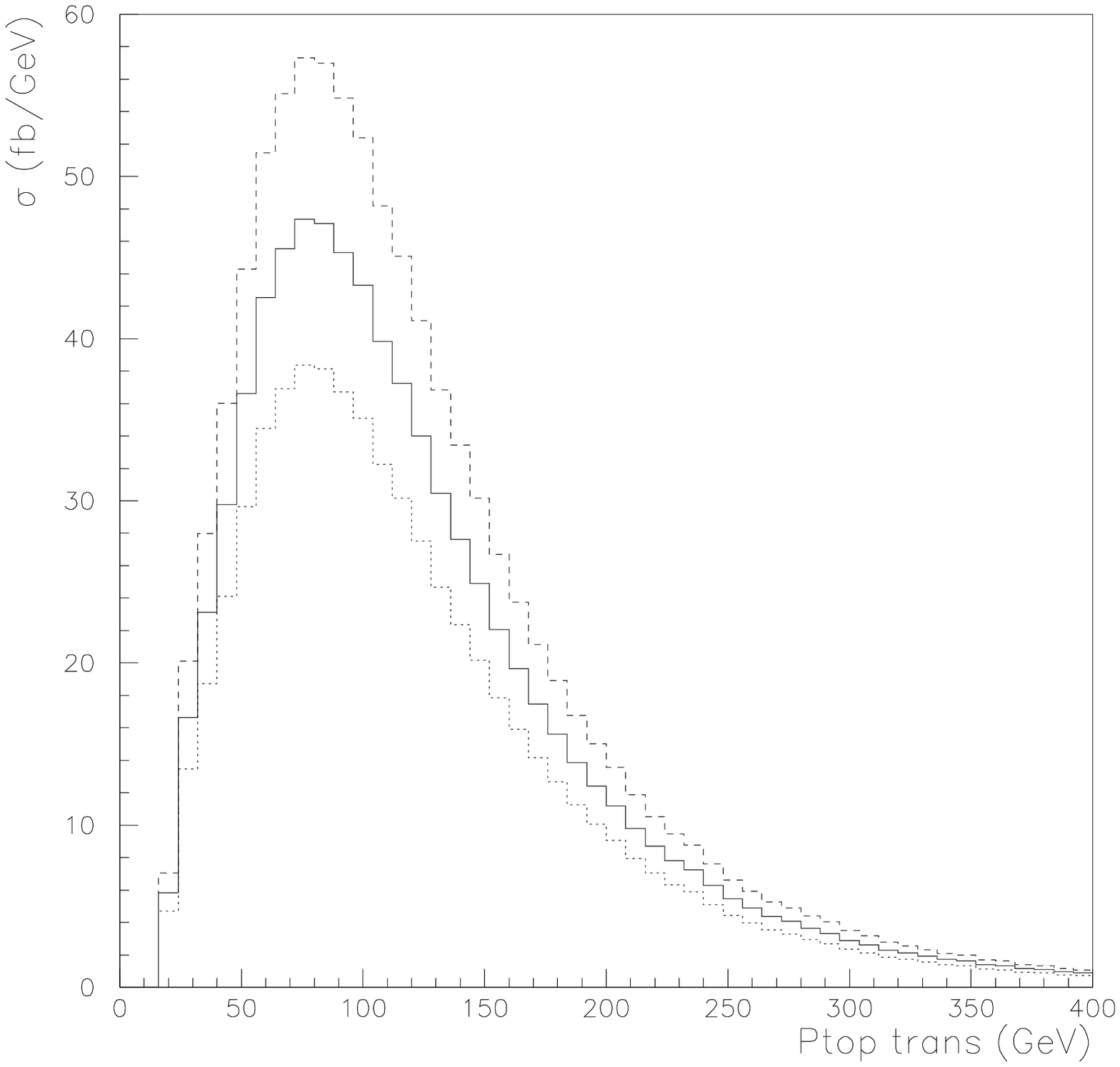}}
\caption{Top transversal momentum distribution corresponding to polarized
single top production at the LHC. All plots correspond to tops polarized in
the spectator jet {\em positive} direction in the {\em top rest frame}. The
subprocesses contributing to the solid line histogram have been calculated
at tree level in the SM ($g_{L}=1$, $g_{R}=0$). The dashed (dotted) line
histogram have been calculated at tree level with $g_{L}=1.1$, and $g_{R}=0$
($g_{L}=0.9$, and $g_{R}=0$). Note the roughly 20 percent variation in this
cross section due to the variation of the left coupling around its SM tree
level value}
\label{ptoptrans_rep_+q2_gl=1+-.1_gr=0}
\end{figure}

\begin{figure}[tbp]
\epsfysize=10.cm 
\centerline{\epsffile{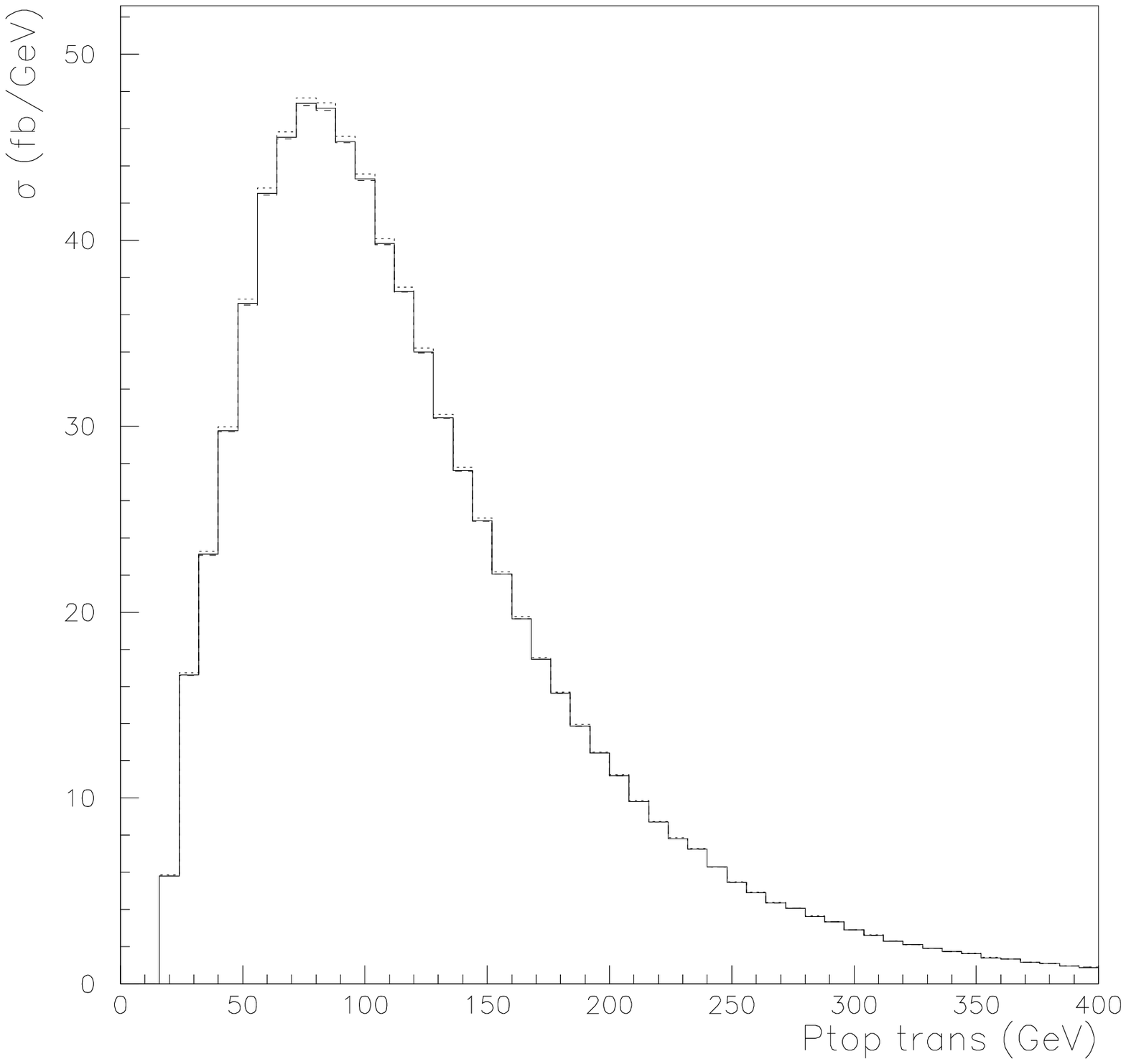}}
\caption{Top transversal momentum distribution corresponding to polarized
single top production at the LHC. All plots correspond to tops polarized in
the spectator jet {\em positive} direction in the {\em top rest frame}. The
subprocesses contributing to the solid line histogram have been calculated
at tree level in the SM ($g_{L}=1$, $g_{R}=0$). The dashed (dotted) line
histogram have been calculated at tree level with $g_{L}=1$, and $g_{R}=0.1$
($g_{L}=1$, and $g_{R}=-0.1$). Note the variation in this cross section due
to the variation of the right coupling around its SM tree level value is
practically inappreciable.}
\label{ptoptrans_rep_+q2_gl=1_gr=+-.1}
\end{figure}

\begin{figure}[tbp]
\epsfysize=10.cm 
\centerline{\epsffile{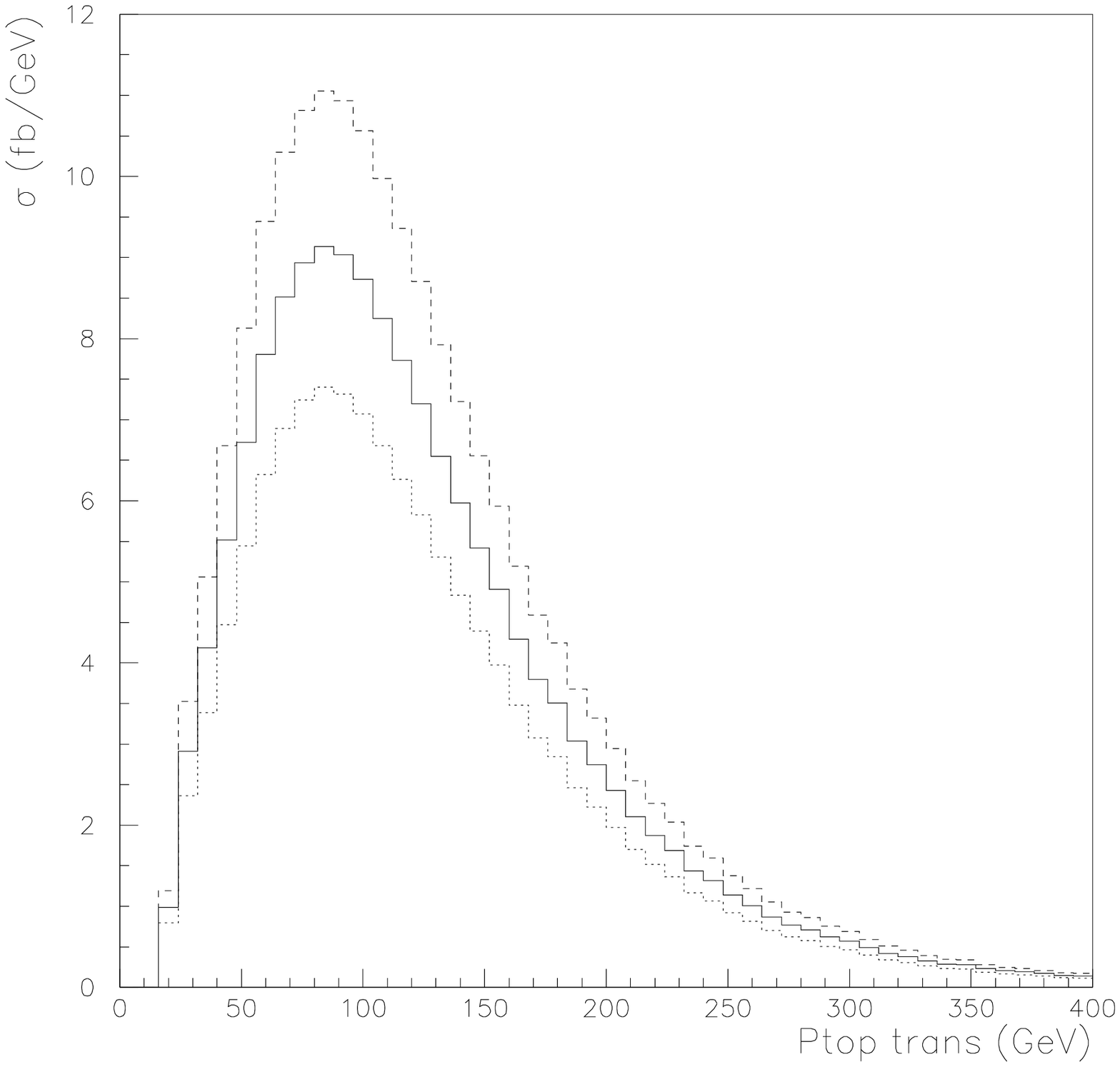}}
\caption{Top transversal momentum distribution corresponding to polarized
single top production at the LHC. All plots correspond to tops polarized in
the spectator jet {\em negative} direction in the {\em top rest frame}. The
subprocesses contributing to the solid line histogram have been calculated
at tree level in the SM ($g_{L}=1$, $g_{R}=0$). The dashed (dotted) line
histogram have been calculated at tree level with $g_{L}=1.1$, and $g_{R}=0$
($g_{L}=0.9$, and $g_{R}=0$).}
\label{ptoptrans_rep_-q2_gl=1+-.1_gr=0}
\end{figure}

\begin{figure}[tbp]
\epsfysize=10.cm 
\centerline{\epsffile{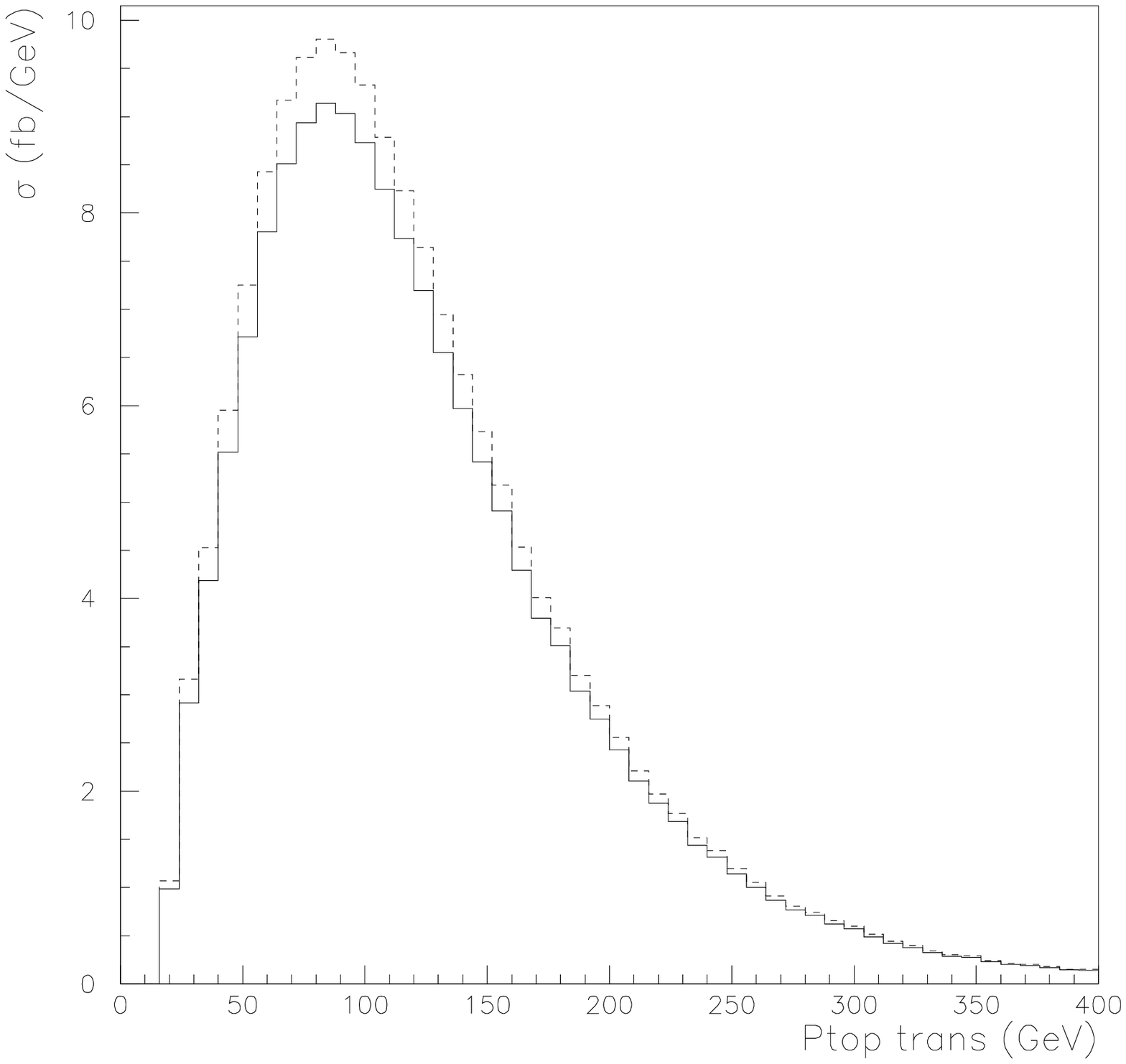}}
\caption{Top transversal momentum distribution corresponding to polarized
single top production at the LHC. All plots correspond to tops polarized in
the spectator jet {\em negative} direction in the {\em top rest frame}. The
subprocesses contributing to the solid line histogram have been calculated
at tree level in the SM ($g_{L}=1$, $g_{R}=0$). The dashed line histogram
have been calculated at tree level with $g_{L}=1$, and $g_{R}=-0.1$. The
histogram corresponding to the couplings $g_{L}=1$, and $g_{R}=0.1$ has been
omitted for the sake of clarity and it is located roughly in between the
ones showed in the picture. Note that the sensitivity of this cross section
to variations of the right coupling around its SM tree level value is
comparable to the sensitivity to variations of the left coupling around its
SM tree level value (see figure \ref{ptoptrans_rep_-q2_gl=1+-.1_gr=0}).}
\label{ptoptrans_rep_-q2_gl=1_gr=+-.1}
\end{figure}

Let us now depart from the tree-level Standard Model and consider non-zero
values for $\delta g_{L}$ and $\delta g_{R}$. . Our results can be
summarized in Figs \ref{ptoptrans_rep_+q2_gl=1+-.1_gr=0}, \ref
{ptoptrans_rep_+q2_gl=1_gr=+-.1}, \ref{ptoptrans_rep_-q2_gl=1+-.1_gr=0}, \ref
{ptoptrans_rep_-q2_gl=1_gr=+-.1}

Taking into account the results of Eq. (\ref{results}) we can establish the
intervals where the effective couplings are indistinguishable from their
tree level Standard Model values taking a 1 sigma deviation as a rough
statistical criterion. Evidently we do not pretend to make here a serious
experimental analysis since we are not taking into account the full set of
experimental and theoretical uncertainties. Our aim is just to present and
order of magnitude estimate of the sensitivity of the different spin basis
to the value of the effective coupling around their tree level Standard
Model value. The results are given in Table \ref{sensitivity}, where we
indicate also the polarization vector chosen in each case. Of course those
sensitivities (which, as said, are merely indicative) are calculated with
the assumption that one could perfectly measure the top polarization in any
of the above basis. As it is well known the top polarization is only
measurable in an indirect way through the angular distribution of its decay
products. In section \ref{decay} we outline the procedure to use our results
to obtain a final angular distribution for the polarized top decay products
(we believe that some confusion exists on this point). Obtaining that
angular distribution involves a convolution of the single top production
cross section with the decay products angular distribution and because of
that we expect the true sensitivity to become worse than the ones given in
table \ref{sensitivity}. Obviously that true sensitivity must be independent
of the spin basis one uses at an intermediate step (see section \ref{decay})
so the whole discussion as to which is the best basis for the top
polarization is academic. However it may still be useful to know that some
basis are more sensitive to the effective couplings than others if one {\em %
assumes} (at least as a gedanken experiment) that the polarization of the
top could be measured directly.

It is worth mentioning that the bottom mass, which appears in the cross
section in crossed left-right terms, such as $m_{b}g_{L}g_{R}$, plays a
crucial role in the actual determination of $g_{R}$. This is because from
the $|{\rm Re}(\kappa _{R}^{CC})|\leq 0.4\times 10^{-2}$ bound \cite{LPY} we
expect $g_{L}g_{R}m_{b}>g_{R}^{2}m_{t}.$ Evidently for the $ts$ or $td$
couplings these terms are not expected to be so relevant.

\begin{table}[tbp]
\centering
\begin{tabular}{|c|c|c|}
\hline
polarization frame & $g_{L}$ & $g_{R}$ \\ \hline
$\hat{n}=\pm \frac{\vec{p}_{1}}{\left| \vec{p}_{1}\right| },~lab$ & $\left[
0.\,\allowbreak 9993,1.\,\allowbreak 0007\right] \qquad \left( -\right) $ & $%
\left[ -0.015,0.074\right] \allowbreak \qquad \left( +\right) $ \\ \hline
$\hat{n}=\pm \frac{\vec{q}_{2}}{\left| \vec{q}_{2}\right| },~lab$ & $\left[
0.9994,1.0006\right] \qquad \left( +\right) $ & $\left[ -0.013,0.054\right]
\allowbreak \qquad \left( -\right) $ \\ \hline
$\hat{n}=\pm \frac{\vec{q}_{2}}{\left| \vec{q}_{2}\right| },~rest$ & $\left[
0.\,\allowbreak 9994,1.\,\allowbreak 0006\right] \qquad \left( +\right) $ & $%
\left[ -0.012,0.051\right] \allowbreak \qquad \left( -\right) $ \\ \hline
\end{tabular}
\caption{Sensitivity of the polarized single top production to variations of
the effective couplings. To calculate the intervals we have taken 1 sigma
statistical deviations from tree level values as an order of magnitude
criterion. Of course, given the uncertainties in the QCD scale, the overall
normalization is dubious and the actual precision on $g_{L}$ a lot less. The
purpose of these figures is to illustrate the relative accuracy. Between
parenthesis we indicate the spin direction taken to calculate each interval.}
\label{sensitivity}
\end{table}

\section{The differential cross section for polarized tops}

We define the matrix element of the subprocess of Figs.\ref{u+gt+b-d+tot},
to \ref{u-gt-b+d-tot} as $M_{+}^{d}$, $M_{+}^{\bar{u}}$, $M_{-}^{u}$, and $%
M_{-}^{\bar{d}}$, respectively. With these definitions the differential
cross section for polarized tops $\sigma $ can be written schematically as 
\[
\sigma =c\left( f_{u}\left| M_{+}^{d}\right| ^{2}+f_{\bar{d}}\left| M_{+}^{%
\bar{u}}\right| ^{2}\right) ,
\]
where $f_{u}$ and $f_{\bar{d}}$ denote the parton distribution functions
corresponding to extracting a $u$-type quark and a $\bar{d}$-type quark
respectively and $c$ is a proportionality factor incorporating the
kinematical and measure factors$.$ Now using our analytical results for the
matrix elements given in the Appendix along with Eq. (\ref{change}) and
symmetries (\ref{sym}) we obtain 
\begin{eqnarray*}
\sigma  &=&cf_{u}\left[ \left| g_{L}\right| ^{2}\left( a+a_{n}\right)
+\left| g_{R}\right| ^{2}\left( b+b_{n}\right) +\frac{g_{R}^{\ast
}g_{L}+g_{R}g_{L}^{\ast }}{2}\left( c+c_{n}\right) +i\frac{g_{L}^{\ast
}g_{R}-g_{R}^{\ast }g_{L}}{2}d_{n}\right]  \\
&&+cf_{\bar{d}}\left[ \left| g_{R}\right| ^{2}\left( a-a_{n}\right) +\left|
g_{L}\right| ^{2}\left( b-b_{n}\right) +\frac{g_{R}^{\ast
}g_{L}+g_{R}g_{L}^{\ast }}{2}\left( c-c_{n}\right) -i\frac{g_{L}^{\ast
}g_{R}-g_{R}^{\ast }g_{L}}{2}d_{n}\right]  \\
&=&\left( 
\begin{array}{cc}
g_{L}^{\ast } & g_{R}^{\ast }
\end{array}
\right) A\left( 
\begin{array}{c}
g_{L} \\ 
g_{R}
\end{array}
\right) ,
\end{eqnarray*}
where 
\begin{equation}
A=c\left( 
\begin{array}{cc}
f_{u}\left( a+a_{n}\right) +f_{\bar{d}}\left( b-b_{n}\right)  & \frac{1}{2}%
f_{u}\left( c+c_{n}+id_{n}\right) +\frac{1}{2}f_{\bar{d}}\left(
c-c_{n}-id_{n}\right)  \\ 
\frac{1}{2}f_{u}\left( c+c_{n}-id_{n}\right) +\frac{1}{2}f_{\bar{d}}\left(
c-c_{n}+id_{n}\right)  & f_{u}\left( b+b_{n}\right) +f_{\bar{d}}\left(
a-a_{n}\right) 
\end{array}
\right) ,  \label{new}
\end{equation}
and where $a,$ $b$, $c$, $a_{n}$, $b_{n}$, $c_{n}$ and $d_{n}$ are
independent of the effective couplings $g_{R}$ and $g_{L}$ and the
subscripts $n$ indicate linear dependence on the top spin four-vector $n.$.
From Eq. (\ref{new}) we observe that $A$ is an Hermitian matrix and
therefore it is diagonalizable with real eigenvalues. Moreover, from the
positivity of $\sigma $ we immediately arrive at the constraints 
\begin{eqnarray}
\det A &\geq &0,  \label{constr1} \\
TrA &\geq &0,  \label{constr2}
\end{eqnarray}
that is 
\begin{eqnarray}
&&\left( f_{u}\left( a+a_{n}\right) +f_{\bar{d}}\left( b-b_{n}\right)
\right) \left( f_{u}\left( b+b_{n}\right) +f_{\bar{d}}\left( a-a_{n}\right)
\right)   \nonumber \\
&\geq &\frac{1}{4}\left( c^{2}\left( f_{u}+f_{\bar{d}}\right) ^{2}+\left(
c_{n}^{2}+d_{n}^{2}\right) \left( f_{u}-f_{\bar{d}}\right)
^{2}+2cc_{n}\left( f_{u}^{2}-f_{\bar{d}}^{2}\right) \right) ,  \label{co1}
\end{eqnarray}
and 
\begin{equation}
\left( f_{u}+f_{\bar{d}}\right) \left( a+b\right) +\left( f_{u}-f_{\bar{d}%
}\right) \left( a_{n}+b_{n}\right) \geq 0.  \label{co2}
\end{equation}
Note that it is not possible to saturate both constraints for the same
configuration because this would imply a vanishing $A$ which in turn would
imply relations such as 
\[
\frac{a+b}{a_{n}+b_{n}}=\frac{f_{\bar{d}}-f_{u}}{f_{\bar{d}}+f_{u}}=\frac{%
a_{n}-b_{n}}{a-b},
\]
which evidently do not hold. Moreover, since constraints (\ref{co1}) and (%
\ref{co2}) must be satisfied for any set of positive PDF's we immediately
obtain the bounds 
\begin{eqnarray*}
ab+a_{n}b_{n}-\frac{1}{4}\left( c^{2}+c_{n}^{2}+d_{n}^{2}\right)  &\geq
&\left| a_{n}b+ab_{n}-\frac{1}{2}cc_{n}\right|  \\
b^{2}+a^{2}-\left( b_{n}^{2}+a_{n}^{2}\right)  &\geq &\frac{1}{2}\left(
c^{2}-\left( c_{n}^{2}+d_{n}^{2}\right) \right) .
\end{eqnarray*}
In order to have a 100\% polarized top we need a spin four-vector $n$ that
saturates the constraint (\ref{constr1}) (that is Eq.(\ref{co1})) for each
kinematical situation, that is we need $A\left( n\right) $ to have a zero
eigenvalue which is equivalent to have a unitary matrix $C$ satisfying 
\[
C^{\dagger }AC={\rm diag}\left( \lambda ,0\right) ,
\]
for some positive eigenvalue $\lambda $. . In general such $n$ need not
exist and, should it exist, is in any case independent of the effective
couplings $g_{R}$ and $g_{L}$. Moreover, provided this $n$ exists there is
only one solution (up to a global complex normalization factor $\alpha $)
for the pair $\left( g_{R},g_{L}\right) $ to the equation $\sigma =0,$ This
solution is just 
\begin{eqnarray}
g_{L} &=&\alpha C_{12},  \nonumber \\
g_{R} &=&\alpha C_{22}.  \label{100}
\end{eqnarray}
Note that if one of the effective couplings vanishes we can take the other
constant and arbitrary. However if both effective couplings are
non-vanishing we would have a quotient $g_{R}/g_{L}$ that would depend in
general on the kinematics. This is not possible so we can conclude that for
a non-vanishing $g_{R}$ ( $g_{L}$ is evidently non-vanishing) it is not
possible to have a pure spin state (or, else, only for fine tuned $g_{R}$ a
100\% polarization is possible).

To illustrate these considerations let us give an example: in the unphysical
situation where $m_{t}\rightarrow 0$ it can be shown that there exists two
solutions to the saturated constraint (\ref{constr1}), namely 
\begin{equation}
m_{t}n^{\mu }\rightarrow \pm \left( \left| \vec{p}_{1}\right| ,p_{1}^{0}%
\frac{\vec{p}_{1}}{\left| \vec{p}_{1}\right| }\right) ,
\end{equation}
once we have found this result we plug it in the expression (\ref{100}) and
we find the solutions $\left( 0,g_{L}\right) $ with $g_{L}$ arbitrary for
the $+$ sign and $\left( g_{R},0\right) $ with $g_{R}$ arbitrary for the $-$
sign. That is, physically we have zero probability of producing a right
handed top when we have only a left handed coupling and viceversa when we
have only a right handed coupling. Note that in this case it is clear that
having both effective couplings non-vanishing would imply the absence of 100
\% polarization in any spin basis. This can be understood in general
remembering that the top particle forms in general an entangled state with
the other particles of the process. Since we are tracing over the unknown
spin degrees of freedom and over the flavors of the spectator quark we do
not expect in general to end up with a top in a pure polarized state;
although this is not impossible as it is shown the in the last example.

In the physical situation where $m_{t}\neq 0$ (we use $m_{t}=175.6$ GeV. and 
$m_{b}=5$ GeV. in this paper) we have found that a spin basis with
relatively high polarization is the one with the spin $\hat{n}$ taken in the
direction of the spectator quark in the top rest frame. This is in
accordance to the results in \cite{mandp}. In general the degree of
polarization ($\frac{N_{+\hat{n}}}{N_{+\hat{n}}+N_{-\hat{n}}}$) depends not
only on the spin frame but also on the particular cuts chosen. We have found
that the lower cut for the transverse momentum of the bottom worsens the
polarization degree but, in spite of that, from Eq.(\ref{results}) we see
that we have a 84\% of polarization in the Standard Model ($g_{L}=1,$ $%
g_{R}=0$) that is much bigger than the 69\% obtained with the helicity
frame. The above results follow the general trend of those presented by
Mahlon and Parke \cite{mandp}, but in general, their degree of polarization
is higher. We understand that this is due to the different cuts (in
particular for the transversal momentum of the bottom) along with the
different set of PDF's used in our simulations.

\section{Measuring the top polarization from its decay products}

\label{decay}A well know result in the tree level SM regarding the measure
of the top polarization from its decay products is the formula that states
the following: Given a top polarized in the $\hat{n}$ direction in its rest
frame, the lepton $l^{+}$ produced in the decay of the top via the process 
\begin{equation}
t\rightarrow b\left( W^{+}\rightarrow l^{+}\nu _{l}\right) ,
\end{equation}
presents an angular distribution \cite{cos}

\begin{equation}
\sigma _{l}=\alpha \left( 1+\cos \theta \right) ,  \label{pol1}
\end{equation}
where $\alpha $ is a normalization factor and $\theta $ is the axial angle
measured from the direction of $\hat{n}.$ What can we do when the top is in
a mixed state with no 100\% polarization in any direction? The first naive
answer would be: With any axis $\hat{n}$ in the top rest frame the top will
have a polarization $p_{+}$ (with $0\leq p_{+}\leq 1)$ in that direction and
a polarization $p_{-}=1-p_{+}$ in the opposite direction so the angular
distribution for the lepton is 
\begin{eqnarray}
\sigma _{l} &=&\alpha \left( p_{+}\left( 1+\cos \theta \right) +p_{-}\left(
1-\cos \theta \right) \right)  \nonumber \\
&=&\alpha \left( 1+\left( p_{+}-p_{-}\right) \cos \theta \right)  \nonumber
\\
&=&\alpha \left( 1+\left( 2p_{+}-1\right) \cos \theta \right) .  \label{pol}
\end{eqnarray}
The problem with formula (\ref{pol}) is that the angular distribution for
the lepton depends on the arbitrary chosen axis $\hat{n}$ and this cannot be
correct. The correct answer can be obtained by noting the following facts:

\begin{itemize}
\item  Given an arbitrary chosen axis $\hat{n}$ in the rest frame and the
associated spin basis to it $\left\{ \left| +\hat{n}\right\rangle ,\left| -%
\hat{n}\right\rangle \right\} $ the top spin state in given by a $2\times 2$
density matrix $\rho $ 
\begin{equation}
\rho =\rho _{+}\left| +\hat{n}\right\rangle \left\langle +\hat{n}\right|
+\rho _{-}\left| -\hat{n}\right\rangle \left\langle -\hat{n}\right| +b\left|
+\hat{n}\right\rangle \left\langle -\hat{n}\right| +b^{\ast }\left| -\hat{n}%
\right\rangle \left\langle +\hat{n}\right| ,
\end{equation}
which is in general not diagonal ($b\neq 0$) and whose coefficients depend
on the rest of kinematical variables determining the differential cross
section.

\item  From the calculation of the polarized cross section {\em we only know}
the diagonal elements $\rho _{\pm }=p_{\pm }=\left| M\right| _{\pm \hat{n}%
}^{2}/\left( \left| M\right| _{+\hat{n}}^{2}+\left| M\right| _{-\hat{n}%
}^{2}\right) .$

\item  Given $\rho $ in any orthogonal basis determined (up to phases) by $%
\hat{n}$ we can change to another basis that diagonalizes $\rho .$ Since the
top is a spin $1/2$ particle, this basis will correspond to another
direction $\hat{n}_{d}$.

\item  Once we have $\rho $ diagonalized then Eq.(\ref{pol}) is trivially
correct with $p_{\pm }=\rho _{\pm }$ and now $\theta $ is unambiguously
measured from the direction of $\hat{n}_{d}.$
\end{itemize}

From the above facts the first question that comes to our minds is if there
exists a way to determine $\hat{n}_{d}$ without knowing the off-diagonal
matrix elements of $\rho .$ The answer is yes. It is an easy exercise of
elementary quantum mechanics that given a $2\times 2$ Hermitian matrix $\rho 
$ the eigenvector with largest (lowest) eigenvalue correspond to the unitary
vector that maximizes (minimizes) the bilinear form $\left\langle v\right|
\rho \left| v\right\rangle $ constrained to $\left\{ \left| v\right\rangle
,~\left\langle v|v\right\rangle =1\right\} $. Since an arbitrary normalized $%
\left| v\right\rangle $ can be written (up to phases) as $\left| +\hat{n}%
\right\rangle $ and in that case $\rho _{+}=p_{+}$ then the correct $\hat{n}%
_{d}$ entering in Eq.(\ref{pol}) is the one that maximizes the differential
cross section $\left| M\right| _{\hat{n}}^{2}$ for each kinematical
configuration. At the end, the correct angular distribution for the leptons
is given by the cross section for polarized tops{\em \ in this basis} ($\hat{%
n}_{d}$) convoluted with formula (\ref{pol1}) (or improvements of it \cite
{Mauser}).

The above analysis was carried out in the Standard Model ($g_{R}=0$) but it
is correct also for $g_{R}\neq 0$ using the complete formula for this case 
\begin{equation}
\sigma _{l}=\alpha \left( 1+\left( p_{+}-p_{-}\right) \cos \theta \left( 1-%
\frac{1}{4}\left| g_{R}\right| ^{2}h\left( \frac{M_{W}^{2}}{m_{t}^{2}}%
\right) \right) \right) ,  \label{final}
\end{equation}
where $h\left( \frac{M_{W}^{2}}{m_{t}^{2}}\right) \simeq 0.566$ \cite{cos}.
Formula (\ref{final}) deserves some comments:

\begin{itemize}
\item  First of all we remember that $\theta $ is the angle (in the top rest
frame) between the $\hat{n}$ that maximizes the difference $\left(
p_{+}-p_{-}\right) $ and the three momentum of the lepton.

\item  Taking into account the above comment and that $\left(
p_{+}-p_{-}\right) $ depends on $g_{L}$ and $g_{R}$ we see that also $\theta 
$ depends on $g_{L}$ and $g_{R}$.

\item  From the computational point of view, formula (\ref{final}) is not an
explicit formula because involves a process of maximization for each
kinematical configuration.

\item  In some works in the literature \cite{mandp} formula (\ref{final}) is
presented for an arbitrary choice of the spin basis $\left\{ \left| \pm \hat{%
n}\right\rangle \right\} $ in the top rest frame. This is incorrect because
it does not take into account that, in general, the top spin density matrix
is not diagonal.

\item  In a recent work \cite{Mauser} $O\left( \alpha _{s}\right) $
corrections are incorporated to the polarized top decay angular analysis. In
this work the density matrix for the top spin is properly taken into
account. To connect this work with ours we have to replace their
polarization vector $\vec{P}$ by $P\hat{n}_{d}$ where the magnitude of the
top polarization $P$ is just the spin asymmetry $\left( \left| M\right| _{+%
\hat{n}_{d}}^{2}-\left| M\right| _{-\hat{n}_{d}}^{2}\right) /\left( \left|
M\right| _{+\hat{n}_{d}}^{2}+\left| M\right| _{-\hat{n}_{d}}^{2}\right) $ in
our language. This taken into account, the density matrix the authors of 
\cite{Mauser} quote is in the basis $\left\{ \left| +\hat{n}%
_{W^{+}}\right\rangle ,\left| -\hat{n}_{W^{+}}\right\rangle \right\} $ where 
$\hat{n}_{W^{+}}$ is a normal vector in the direction of the three momentum
of the $W^{+}$ (in the top rest frame).
\end{itemize}

\section{Conclusions}

We have done a complete calculation of the subprocess cross sections for
polarized tops or anti-tops including the right effective coupling and
bottom mass corrections. We have used a $p_{T}>30$ GeV. cut in the
transverse momentum of the produced $\bar{b}$ quark and, accordingly we have
retained only the so called $2\rightarrow 3$ process, for the reasons
described in the text.

Our analysis here is completely general. No approximation is made. We use
the most general set of couplings and, since our approach is completely
analytical, we can describe the contribution from other intermediate quarks
in the $t$ channel, mixing, etc. Masses and mixing angles are retained. On
the contrary, the analysis has to be considered only preliminary from an
experimental point of view. No detailed study of the backgrounds has been
made, except for the dominant $gg\rightarrow t\bar{t}$ process which has
been considered to some extent (although again without quantitative
evaluation).

Given the (presumed) smallness of the right handed couplings, the bottom
mass plays a role which is more important than anticipated, as the mixed
crossed $g_{L}g_{R}$ term, which actually is the most sensitive one to $%
g_{R} $ is accompanied by a $b$ quark mass. The statistical sensitivity to
different values of this coupling is given in the text.

We present a variety of $p_{T}$ and angular distributions both for the $t$
and the $\bar{b}$ quarks. Obviously, the top decays shortly after
production, but we have not made detailed simulations of this part. In fact,
the interest of this decay is obvious: one can measure the spin of the top
through the angular distribution of the leptons produced in this decay. In
the Standard Model, single top production gives a high degree of
polarization (84 \% in the optimal basis, with the present set of cuts).
This is a high degree of polarization, but well below the 90+ claimed by
Mahlon and Parke in \cite{mandp}. We understand this being due to the
presence of the 30 GeV. cut. In fact, if we remove this cut completely we
get a 91 \% polarization. Still below the result of \cite{mandp} but in
rough agreement (note that we do not include the $2\rightarrow 2$ process).
Inasmuch as they can be compared our results are in good agreement with
those presented in \cite{SSW} in what concerns the total cross-section. Two
different choices for the strong scale $\mu ^{2}$ are presented.

In addition, it turns out that when $g_{R}\neq 0$ the top can never be 100\%
polarized. In other words, it is in a mixed state. In this case we show that
a unique spin basis is singled out which allows one to connect top decay
products angular distribution with the polarized top differential cross
section.

Finally it should be mentioned that a previous study for this process in the
present context was performed by the present authors\cite{effW} using the
effective W approximation\cite{dawson}, in which the $W$ is treated as a
parton of the proton. While this is certainly not an exact treatment, it is
expected to be sufficiently good for our purposes. In the course of this
work we have found, however, a number of interesting differences. A detailed
comparison with this practical and widely used approach will be given
elsewhere.

\section{Acknowledgments}

We would like to thank A.Dobado, S.Forte, M.J.Herrero, J.R.Pel\'aez and
E.Ru\'\i z-Morales for multiple discussions. We would also like to thank
W.Hollik for encouragement and for allowing us to present preliminary
results in the LHC CERN Workshop. J.M. acknowledges a fellowship from
Generalitat de Catalunya, grant 1998FI-00614. Financial support from grants
AEN98-0431, 1998SGR 00026 and EURODAPHNE is greatly appreciated.

\appendix

\section{Subprocesses cross sections}

\begin{figure}[tbp]
\epsfysize=9.0cm 
\centerline{\epsffile{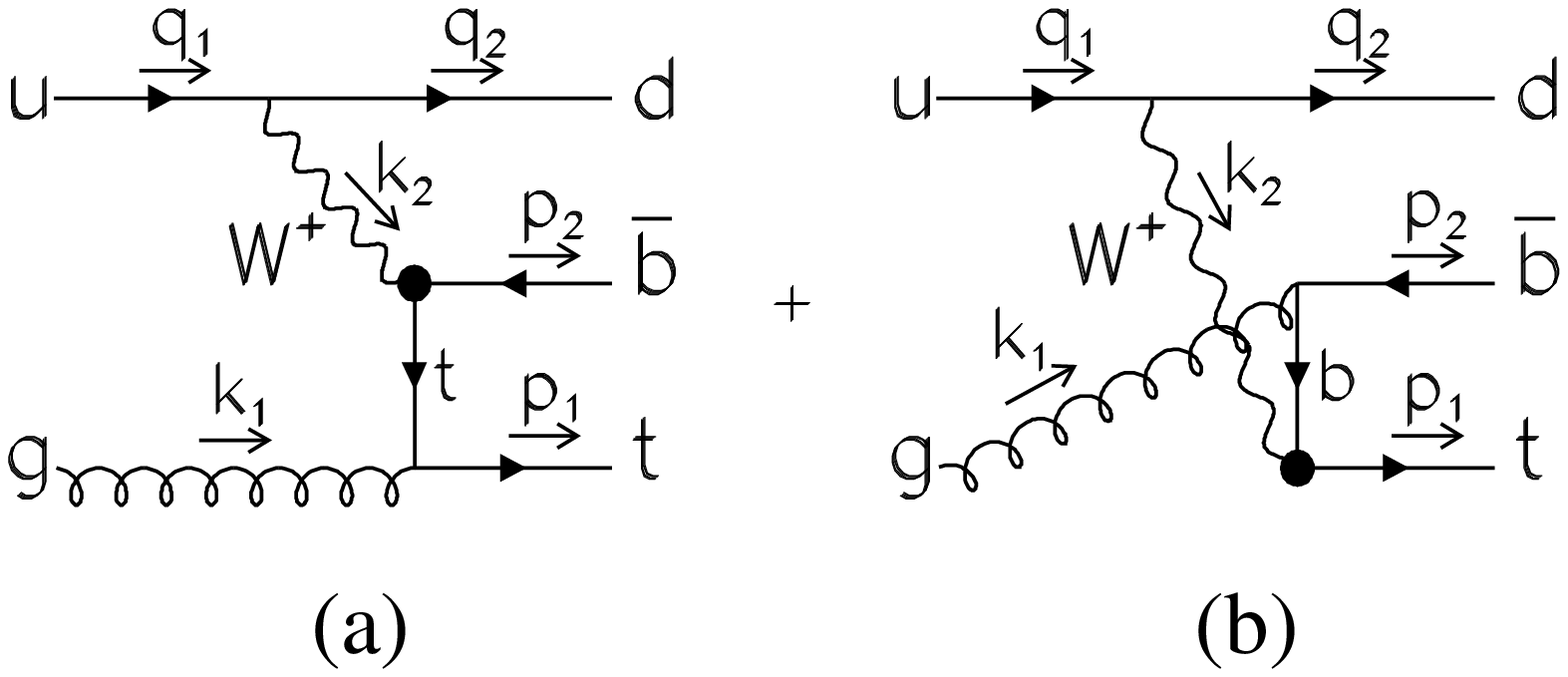}}
\caption{Feynman diagrams contributing single top production subprocess. In
this case we have a $d$ as spectator quark}
\label{u+gt+b-d+tot}
\end{figure}

\begin{figure}[tbp]
\epsfysize=9.0cm 
\centerline{\epsffile{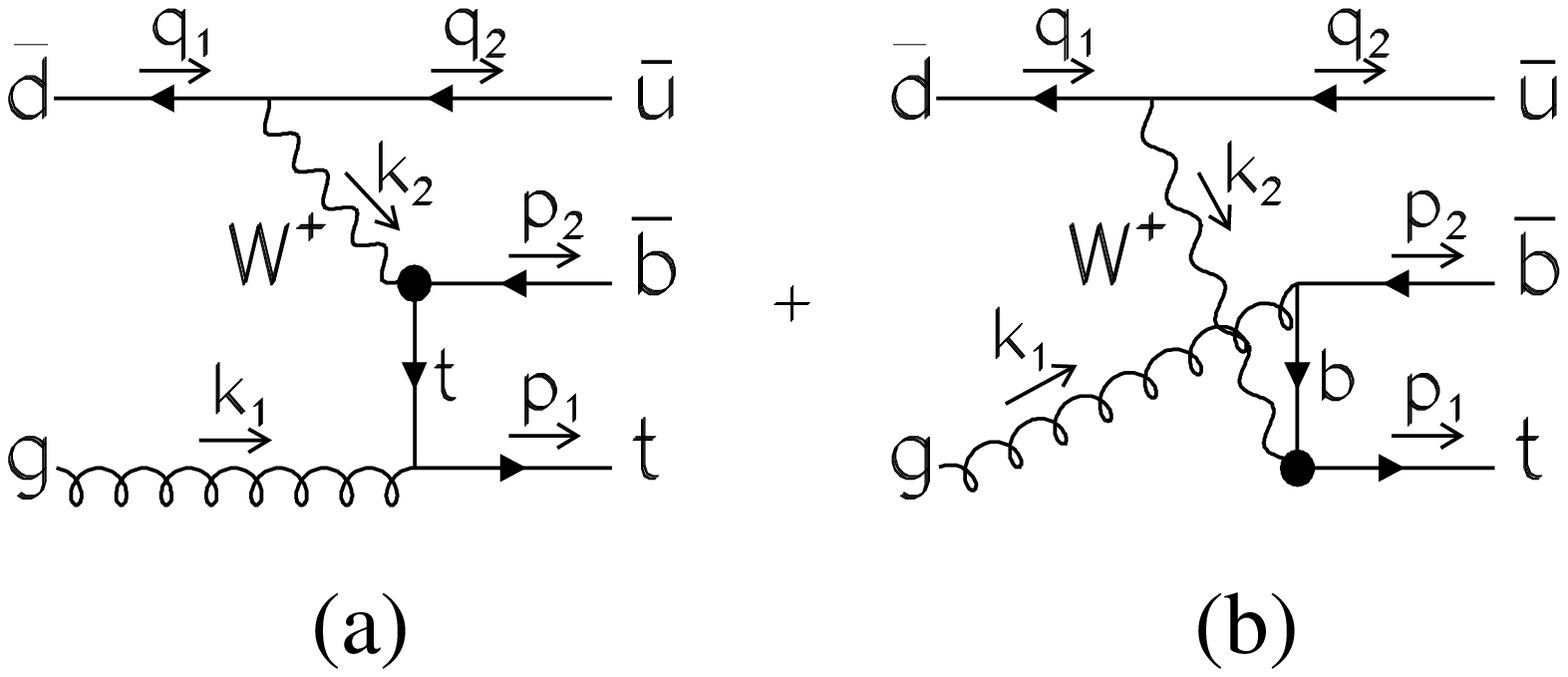}}
\caption{Feynman diagrams contributing single top production subprocess. In
this case we have a $\bar{u}$ as spectator quark}
\label{d-gt+b-u-tot}
\end{figure}

\begin{figure}[tbp]
\epsfysize=9.0cm 
\centerline{\epsffile{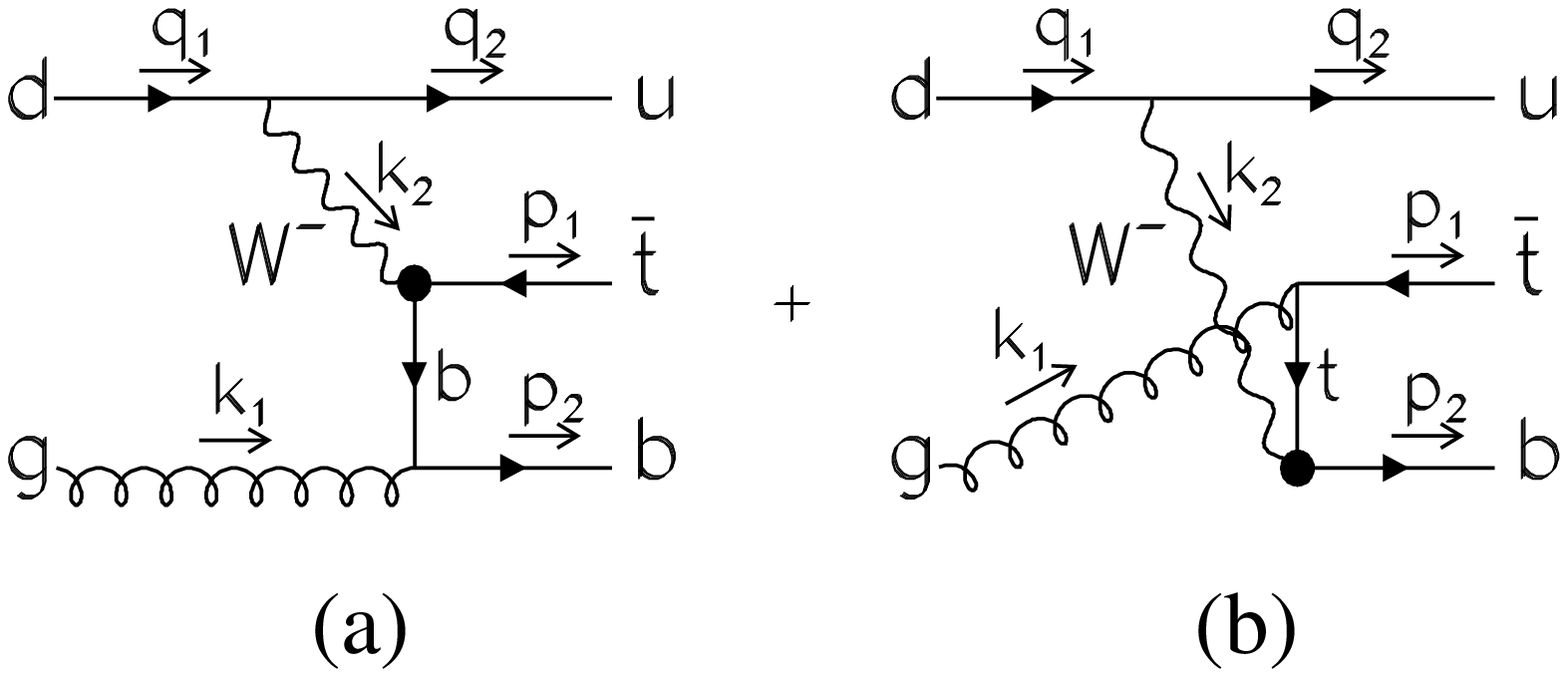}}
\caption{Feynman diagrams contributing single anti-top production
subprocess. In this case we have a $u$ as spectator quark}
\label{d+gt-b+u+tot}
\end{figure}

\begin{figure}[tbp]
\epsfysize=9.0cm 
\centerline{\epsffile{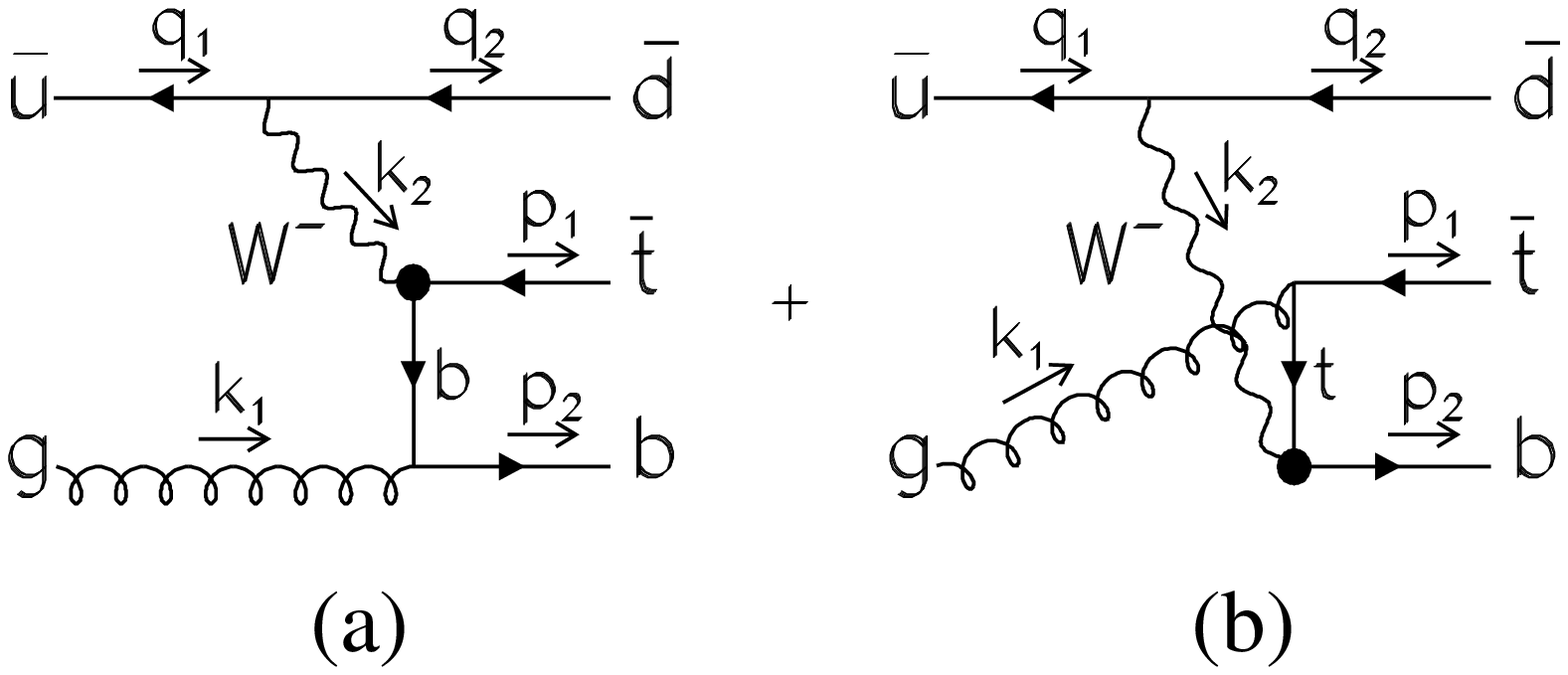}}
\caption{Feynman diagrams contributing single anti-top production
subprocess. In this case we have a $\bar{d}$ as spectator quark}
\label{u-gt-b+d-tot}
\end{figure}

In this appendix we present the analytical results obtained for the matrix
elements corresponding to the processes of Figs.(\ref{u+gt+b-d+tot}-\ref
{u-gt-b+d-tot}) as $M_{+}^{d}$, $M_{+}^{\bar{u}}$, $M_{-}^{u}$, and $M_{-}^{%
\bar{d}}$, respectively. Defining 
\begin{eqnarray*}
g_{+} &=&g_{R}, \\
g_{-} &=&g_{L},
\end{eqnarray*}
we have the square modulus 
\begin{equation}
\left| M_{-}^{u}\right| ^{2}=g_{s}^{2}\left(
O_{11}A_{11}+O_{22}A_{22}+O_{c}\left( A_{p}^{\left( +\right) }+A_{p}^{\left(
-\right) }+A_{m_{t}}^{\left( +\right) }+A_{m_{t}}^{\left( -\right)
}+A_{m_{b}}^{\left( +\right) }+A_{m_{b}}^{\left( -\right) }\right) \right) ,
\label{casifinal}
\end{equation}
with 
\begin{eqnarray}
O_{11} &=&\frac{1}{4\left( k_{1}\cdot p_{1}\right) ^{2}},  \nonumber \\
O_{22} &=&\frac{1}{4\left( k_{1}\cdot p_{2}\right) ^{2}},  \nonumber \\
O_{c} &=&\frac{1}{4\left( k_{1}\cdot p_{1}\right) \left( k_{1}\cdot
p_{2}\right) },  \label{oij}
\end{eqnarray}
and 
\begin{eqnarray*}
A_{11} &=&\frac{\left| g\right| ^{4}\left| K_{ud}\right| ^{2}}{\left(
k_{2}^{2}-M_{W}^{2}\right) ^{2}}\left\{ im_{t}^{2}m_{b}\frac{g_{L}^{\ast
}g_{R}-g_{R}^{\ast }g_{L}}{2}\varepsilon ^{\mu \nu \alpha \beta }\left(
k_{1}-p_{1}\right) _{\mu }n_{\nu }q_{2\alpha }q_{1\beta }\right. \\
&&+m_{t}m_{b}\frac{g_{R}^{\ast }g_{L}+g_{L}^{\ast }g_{R}}{2}\left[
m_{t}\left( q_{2}\cdot \left( k_{1}-p_{1}\right) \right) \left( q_{1}\cdot
n\right) \right. \\
&&-\left. m_{t}\left( q_{1}\cdot \left( k_{1}-p_{1}\right) \right) \left(
q_{2}\cdot n\right) -\left( q_{1}\cdot q_{2}\right) \left( m_{t}^{2}-\left(
k_{1}\cdot p_{1}\right) \right) \right] \\
&&+2\left| g_{L}\right| ^{2}\left( q_{2}\cdot p_{2}\right) \left[ \left(
m_{t}^{2}+\frac{p_{1}+m_{t}n}{2}\cdot \left( k_{1}-p_{1}\right) \right)
\left( q_{1}\cdot \left( k_{1}-p_{1}\right) \right) \right. \\
&&-\left. \frac{1}{2}m_{t}^{3}\left( n\cdot q_{1}\right) +\left( \frac{%
p_{1}+m_{t}n}{2}\cdot q_{1}\right) \left( k_{1}\cdot p_{1}\right) \right] \\
&&+2\left| g_{R}\right| ^{2}\left( q_{1}\cdot p_{2}\right) \left[ \left(
m_{t}^{2}+\frac{p_{1}-m_{t}n}{2}\cdot \left( k_{1}-p_{1}\right) \right)
\left( q_{2}\cdot \left( k_{1}-p_{1}\right) \right) \right. \\
&&+\left. \left. \frac{1}{2}m_{t}^{3}\left( n\cdot q_{2}\right) +\left( 
\frac{p_{1}-m_{t}n}{2}\cdot q_{2}\right) \left( k_{1}\cdot p_{1}\right) %
\right] \right\} ,
\end{eqnarray*}
and 
\begin{eqnarray*}
A_{22} &=&\frac{\left| g\right| ^{4}\left| K_{ud}\right| ^{2}}{\left(
k_{2}^{2}-M_{W}^{2}\right) ^{2}}\left\{ \left( k_{1}\cdot p_{2}\right) \left[
2\left| g_{R}\right| ^{2}\left( q_{1}\cdot k_{1}\right) \left( q_{2}\cdot 
\frac{p_{1}-m_{t}n}{2}\right) \right. \right. \\
&&+\left. 2\left| g_{L}\right| ^{2}\left( q_{2}\cdot k_{1}\right) \left(
q_{1}\cdot \frac{p_{1}+m_{t}n}{2}\right) \right] \\
&&+m_{b}^{2}\left[ 2\left| g_{R}\right| ^{2}\left( q_{1}\cdot \left(
k_{1}-p_{2}\right) \right) \left( q_{2}\cdot \frac{p_{1}-m_{t}n}{2}\right)
\right. \\
&&+\left. 2\left| g_{L}\right| ^{2}\left( q_{2}\cdot \left(
k_{1}-p_{2}\right) \right) \left( q_{1}\cdot \frac{p_{1}+m_{t}n}{2}\right) %
\right] \\
&&+m_{b}\frac{g_{L}^{\ast }g_{R}+g_{R}^{\ast }g_{L}}{2}\left(
m_{b}^{2}-\left( k_{1}\cdot p_{2}\right) \right) \left[ -m_{t}\left(
q_{1}\cdot q_{2}\right) \right. \\
&&-\left. \left( q_{1}\cdot n\right) \left( q_{2}\cdot p_{1}\right) +\left(
q_{2}\cdot n\right) \left( q_{1}\cdot p_{1}\right) \right] \\
&&-\left. im_{b}\frac{g_{L}^{\ast }g_{R}-g_{R}^{\ast }g_{L}}{2}\left(
m_{b}^{2}-\left( k_{1}\cdot p_{2}\right) \right) \varepsilon ^{\mu \nu
\alpha \beta }n_{\mu }p_{1\nu }q_{2\alpha }q_{1\beta }\right\} ,
\end{eqnarray*}
and 
\begin{eqnarray*}
A_{p}^{\left( \pm \right) } &=&-\frac{\left| g\right| ^{4}\left|
K_{ud}\right| ^{2}}{\left( k_{2}^{2}-M_{W}^{2}\right) ^{2}}\left| g_{\pm
}\right| ^{2}\left\{ \left( q_{1}\cdot q_{2}\right) \left[ \left( \left(
k_{1}-p_{1}\right) \cdot \left( k_{2}-p_{1}\right) \right) \left( \frac{%
p_{1}\mp m_{t}n}{2}\cdot p_{2}\right) \right. \right. \\
&&+\left. \left( \left( k_{1}-p_{1}\right) \cdot \frac{p_{1}\mp m_{t}n}{2}%
\right) \left( \left( k_{2}-p_{1}\right) \cdot p_{2}\right) -\left( \left(
k_{1}-p_{1}\right) \cdot p_{2}\right) \left( \frac{p_{1}\mp m_{t}n}{2}\cdot
\left( k_{2}-p_{1}\right) \right) \right] \\
&&+\left( \left( k_{2}-p_{1}\right) \cdot q_{2}\right) \left[ \left(
p_{2}\cdot \left( k_{1}-p_{1}\right) \right) \left( q_{1}\cdot \frac{%
p_{1}\mp m_{t}n}{2}\right) -\left( q_{1}\cdot p_{2}\right) \left( \left(
k_{1}-p_{1}\right) \cdot \frac{p_{1}\mp m_{t}n}{2}\right) \right] \\
&&-\left( \left( k_{1}-p_{1}\right) \cdot q_{2}\right) \left[ \left(
p_{2}\cdot \left( k_{2}-p_{1}\right) \right) \left( q_{1}\cdot \frac{%
p_{1}\mp m_{t}n}{2}\right) -\left( q_{1}\cdot p_{2}\right) \left( \left(
k_{2}-p_{1}\right) \cdot \frac{p_{1}\mp m_{t}n}{2}\right) \right] \\
&&+\left( \left( k_{2}-p_{1}\right) \cdot q_{1}\right) \left[ \left(
p_{2}\cdot \left( k_{1}-p_{1}\right) \right) \left( q_{2}\cdot \frac{%
p_{1}\mp m_{t}n}{2}\right) -\left( q_{2}\cdot p_{2}\right) \left( \left(
k_{1}-p_{1}\right) \cdot \frac{p_{1}\mp m_{t}n}{2}\right) \right] \\
&&-\left( \left( k_{1}-p_{1}\right) \cdot q_{1}\right) \left[ \left(
p_{2}\cdot \left( k_{2}-p_{1}\right) \right) \left( q_{2}\cdot \frac{%
p_{1}\mp m_{t}n}{2}\right) -\left( q_{2}\cdot p_{2}\right) \left( \left(
k_{2}-p_{1}\right) \cdot \frac{p_{1}\mp m_{t}n}{2}\right) \right] \\
&&\pm \left( \left( k_{1}-p_{1}\right) \cdot \left( k_{2}-p_{1}\right)
\right) \left[ \left( \left( \frac{p_{1}\mp m_{t}n}{2}\right) \cdot
q_{2}\right) \left( p_{2}\cdot q_{1}\right) -\left( \left( \frac{p_{1}\mp
m_{t}n}{2}\right) \cdot q_{1}\right) \left( p_{2}\cdot q_{2}\right) \right]
\\
&&\pm \left. \left( \frac{p_{1}\mp m_{t}n}{2}\cdot p_{2}\right) \left[
\left( \left( k_{1}-p_{1}\right) \cdot q_{2}\right) \left( \left(
k_{2}-p_{1}\right) \cdot q_{1}\right) -\left( \left( k_{1}-p_{1}\right)
\cdot q_{1}\right) \left( \left( k_{2}-p_{1}\right) \cdot q_{2}\right) %
\right] \right\} ,
\end{eqnarray*}
and 
\begin{eqnarray*}
A_{m_{t}}^{\left( \pm \right) } &=&\frac{\left| g\right| ^{4}\left|
K_{ud}\right| ^{2}}{\left( k_{2}^{2}-M_{W}^{2}\right) ^{2}}\frac{\left|
g_{\pm }\right| ^{2}}{2}\left\{ \left( m_{t}n\cdot p_{2}\right) \left[
\left( p_{1}\cdot q_{2}\right) \left( \left( k_{2}-p_{1}\right) \cdot
q_{1}\right) -\left( \left( k_{2}-p_{1}\right) \cdot q_{2}\right) \left(
p_{1}\cdot q_{1}\right) \right] \right. \\
&&-\left( m_{t}n\cdot q_{2}\right) \left[ \left( p_{1}\cdot p_{2}\right)
\left( \left( k_{2}-p_{1}\right) \cdot q_{1}\right) -\left( \left(
k_{2}-p_{1}\right) \cdot p_{2}\right) \left( p_{1}\cdot q_{1}\right) \right]
\\
&&+\left( m_{t}n\cdot q_{1}\right) \left[ \left( p_{1}\cdot p_{2}\right)
\left( \left( k_{2}-p_{1}\right) \cdot q_{2}\right) -\left( \left(
k_{2}-p_{1}\right) \cdot p_{2}\right) \left( p_{1}\cdot q_{2}\right) \right]
\\
&&+m_{t}^{2}\left[ \left( q_{2}\cdot p_{2}\right) \left( q_{1}\cdot \left(
k_{2}-p_{1}\right) \right) +\left( q_{1}\cdot p_{2}\right) \left( q_{2}\cdot
\left( k_{2}-p_{1}\right) \right) -\left( q_{1}\cdot q_{2}\right) \left(
p_{2}\cdot \left( k_{2}-p_{1}\right) \right) \right] \\
&&\pm m_{t}\left( n\cdot \left( k_{2}-p_{1}\right) \right) \left[ \left(
q_{2}\cdot p_{2}\right) \left( q_{1}\cdot p_{1}\right) +\left( q_{1}\cdot
p_{2}\right) \left( q_{2}\cdot p_{1}\right) -\left( q_{1}\cdot q_{2}\right)
\left( p_{2}\cdot p_{1}\right) \right] \\
&&\mp \left. m_{t}\left( p_{1}\cdot \left( k_{2}-p_{1}\right) \right) \left[
\left( q_{2}\cdot p_{2}\right) \left( q_{1}\cdot n\right) +\left( q_{1}\cdot
p_{2}\right) \left( q_{2}\cdot n\right) -\left( q_{1}\cdot q_{2}\right)
\left( p_{2}\cdot n\right) \right] \right\} ,
\end{eqnarray*}
and 
\begin{eqnarray*}
A_{m_{b}}^{\left( \pm \right) } &=&\frac{m_{b}\left| g\right| ^{4}\left|
K_{ud}\right| ^{2}}{\left( k_{2}^{2}-M_{W}^{2}\right) ^{2}}\frac{g_{\pm
}^{\ast }g_{\mp }}{2}\left\{ 2\left( p_{1}\cdot p_{2}\right) \left[ \left(
n\cdot q_{2}\right) \left( \left( k_{1}-p_{1}\right) \cdot q_{1}\right)
-\left( n\cdot q_{1}\right) \left( \left( k_{1}-p_{1}\right) \cdot
q_{2}\right) \right] \right. \\
&&-2\left( n\cdot p_{2}\right) \left[ \left( p_{1}\cdot q_{2}\right) \left(
\left( k_{1}-p_{1}\right) \cdot q_{1}\right) -\left( p_{1}\cdot q_{1}\right)
\left( \left( k_{1}-p_{1}\right) \cdot q_{2}\right) \right] \\
&&\pm i\varepsilon ^{\mu \nu \alpha \beta }q_{2\alpha }q_{1\beta }\left(
n_{\mu }p_{1\nu }\left( k_{1}-p_{1}\right) \cdot p_{2}+p_{2\mu }n_{\nu
}\left( k_{1}-p_{1}\right) \cdot p_{1}+p_{1\mu }p_{2\nu }\left(
k_{1}-p_{1}\right) \cdot n\right) \\
&&\mp i\varepsilon ^{\mu \nu \alpha \beta }q_{2\alpha }q_{1\beta }\left(
k_{1}-p_{1}\right) _{\mu }\left[ n_{\nu }\left( p_{1}\cdot \left(
k_{2}-p_{1}\right) \right) +\left( k_{2}-p_{1}\right) _{\nu }\left(
p_{1}\cdot n\right) \right] \\
&&+\left( n\cdot \left( k_{1}-p_{1}\right) \right) \left[ \left( p_{1}\cdot
q_{2}\right) \left( \left( k_{2}-p_{1}\right) \cdot q_{1}\right) -\left(
\left( k_{2}-p_{1}\right) \cdot q_{2}\right) \left( p_{1}\cdot q_{1}\right) %
\right] \\
&&-\left( n\cdot q_{2}\right) \left[ \left( p_{1}\cdot \left(
k_{1}-p_{1}\right) \right) \left( \left( k_{2}-p_{1}\right) \cdot
q_{1}\right) -\left( \left( k_{2}-p_{1}\right) \cdot \left(
k_{1}-p_{1}\right) \right) \left( p_{1}\cdot q_{1}\right) \right] \\
&&+\left( n\cdot q_{1}\right) \left[ \left( p_{1}\cdot \left(
k_{1}-p_{1}\right) \right) \left( \left( k_{2}-p_{1}\right) \cdot
q_{2}\right) -\left( \left( k_{2}-p_{1}\right) \cdot \left(
k_{1}-p_{1}\right) \right) \left( p_{1}\cdot q_{2}\right) \right] \\
&&+2m_{t}\left[ \left( q_{2}\cdot \left( k_{1}-p_{1}\right) \right) \left(
q_{1}\cdot p_{2}\right) +\left( q_{1}\cdot \left( k_{1}-p_{1}\right) \right)
\left( q_{2}\cdot p_{2}\right) -\left( q_{1}\cdot k_{1}\right) \left(
q_{2}\cdot k_{1}\right) \right] \\
&&+\left. m_{t}\left( q_{1}\cdot q_{2}\right) \left[ \left( p_{2}\cdot
p_{1}\right) +\left( \left( k_{1}-p_{1}\right) \cdot \left(
k_{1}-p_{2}\right) \right) \right] \right\} \\
&&+m_{b}^{2}\frac{\left| g_{\pm }\right| ^{2}}{2}\frac{\left| g\right|
^{4}\left| K_{ud}\right| ^{2}\left| K_{tb}\right| ^{2}}{\left(
k_{2}^{2}-M_{W}^{2}\right) ^{2}}\left\{ -m_{t}\left[ \left( n\cdot
q_{2}\right) \left( p_{1}\cdot q_{1}\right) -\left( n\cdot q_{1}\right)
\left( p_{1}\cdot q_{2}\right) \right] \right. \\
&&+m_{t}^{2}\left( q_{1}\cdot q_{2}\right) -2\left[ \left( q_{2}\cdot \left(
k_{1}-p_{1}\right) \right) \left( q_{1}\cdot \frac{p_{1}\mp m_{t}n}{2}%
\right) \right. \\
&&+\left. \left. \left( q_{1}\cdot \left( k_{1}-p_{1}\right) \right) \left(
q_{2}\cdot \frac{p_{1}\mp m_{t}n}{2}\right) -\left( q_{1}\cdot q_{2}\right)
\left( \left( k_{1}-p_{1}\right) \cdot \frac{p_{1}\mp m_{t}n}{2}\right) %
\right] \right\} ,
\end{eqnarray*}
Finally, it can be shown that we can obtain the other matrix elements from
the above expressions performing the following changes 
\begin{equation}
\begin{array}{lcr}
\left| M_{-}^{u}\right| ^{2}\longleftrightarrow \left| M_{+}^{\bar{u}%
}\right| ^{2} & \quad \Leftrightarrow \quad & n\longleftrightarrow -n, \\ 
\left| M_{-}^{u}\right| ^{2}\longleftrightarrow \left| M_{+}^{d}\right| ^{2}
& \quad \Leftrightarrow \quad & g_{L}\leftrightarrow g_{R}^{\ast }, \\ 
\left| M_{-}^{u}\right| ^{2}\longleftrightarrow \left| M_{-}^{\bar{d}%
}\right| ^{2} & \quad \Leftrightarrow \quad & q_{1}\leftrightarrow q_{2},
\end{array}
\label{change}
\end{equation}
it is useful to note also that all matrix elements are symmetric under the
change 
\begin{equation}
\left( n,g_{L},q_{1}\right) \leftrightarrow \left( -n,g_{R}^{\ast
},q_{2}\right) ,  \label{sym}
\end{equation}

\begin{figure}[tbp]
\epsfysize=10.cm 
\centerline{\epsffile{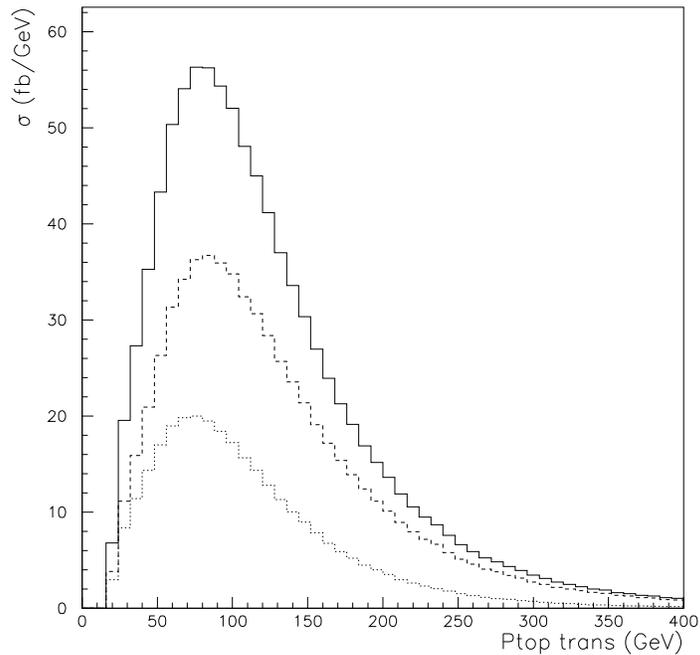}}
\caption{Top transversal momentum distribution corresponding to polarized
single top production at the LHC in the LAB system. The solid line
corresponds to unpolarized top production and the dashed (dotted) line
corresponds to tops of negative (positive) helicity. The subprocesses
contributing to these histograms have been calculated at tree level in the
electroweak theory. The cuts are described in the text. The degree of
polarization in this spin basis and reference frame is only 69\% . The QCD
scale is taken to be $\protect\mu ^{2}=\hat{s}=\left( q_{1}+q_{2}\right)
^{2} $. }
\label{ptoptrans_lab_lr_gl=1_gr=0}
\end{figure}

\begin{figure}[tbp]
\epsfysize=10.cm 
\centerline{\epsffile{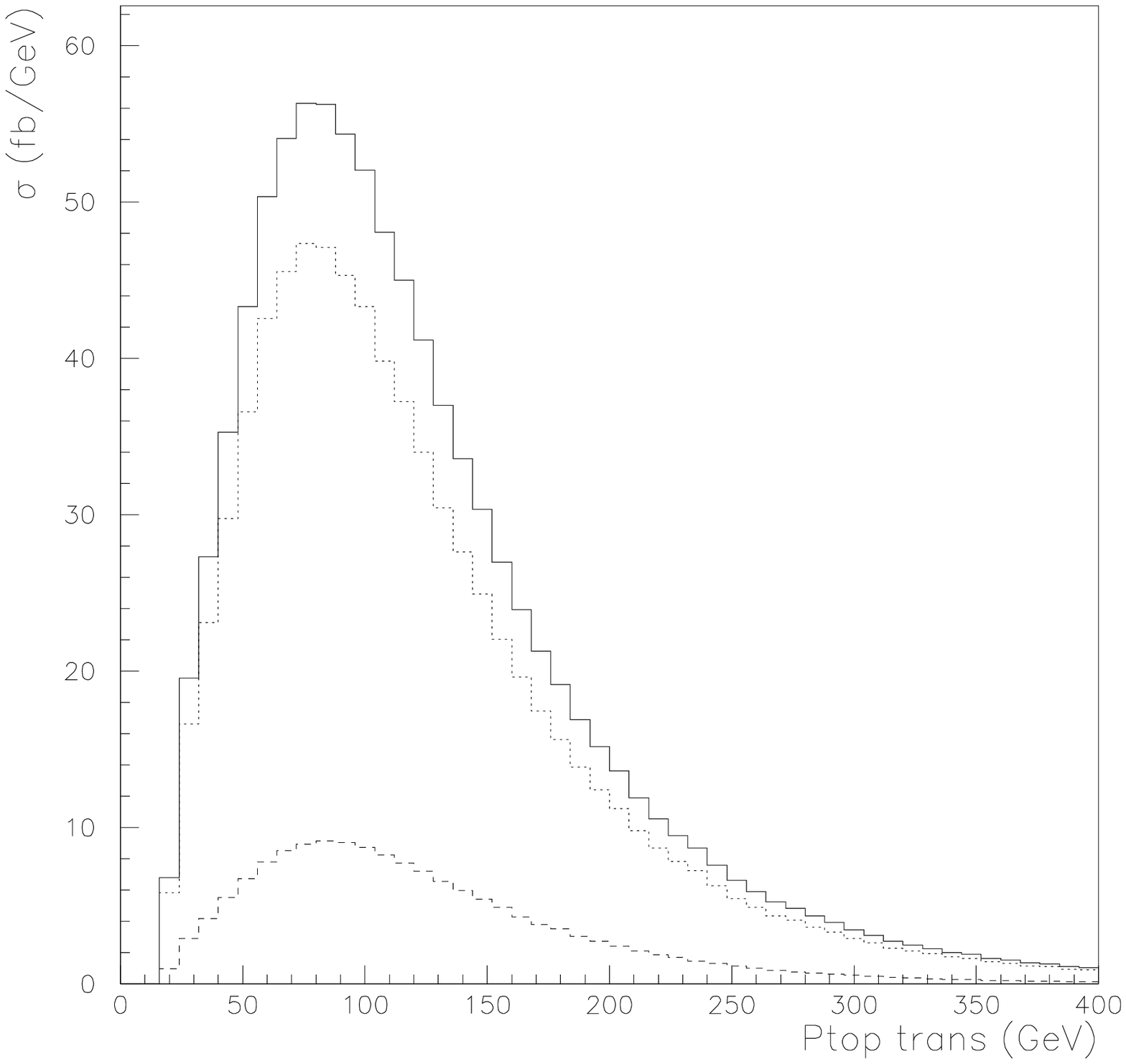}}
\caption{Top transversal momentum distribution corresponding to polarized
single top production at the LHC. The solid line corresponds to unpolarized
top production and the dashed (dotted) line corresponds to tops polarized in
the spectator jet negative (positive) direction in the top rest frame. The
subprocesses contributing to these histograms have been calculated at tree
level in the electroweak theory. With our set of cuts, the polarization is
84 \%. The QCD scale is $\protect\mu ^{2}=\hat{s}=\left( q_{1}+q_{2}\right)
^{2}$ }
\label{ptoptrans_rep_+-q2_gl=1_gr=0}
\end{figure}

\begin{figure}[tbp]
\epsfysize=10.cm 
\centerline{\epsffile{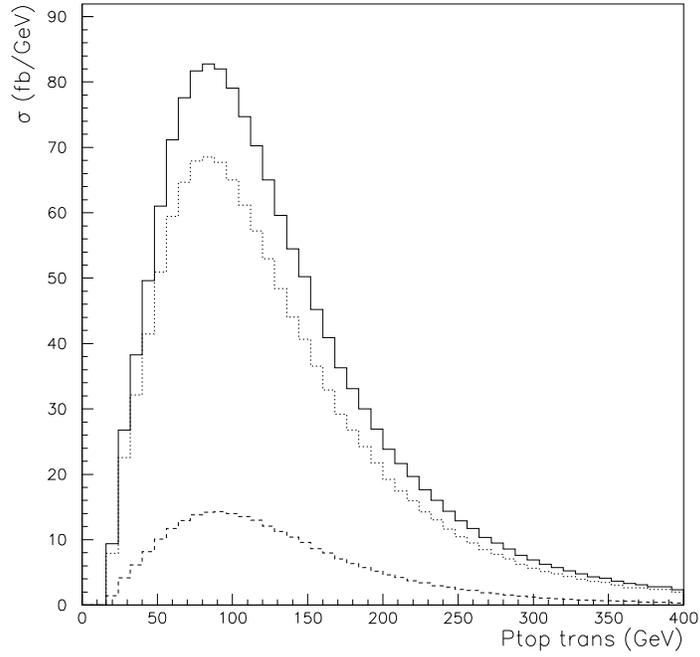}}
\caption{Here we plot the same histograms that in Fig.\ref
{ptoptrans_rep_+-q2_gl=1_gr=0}. However here we have set the QCD scale $%
\protect\mu =p_{cut}^{T(bot)}=30$ GeV. For the gluon PDF we have also used
this scale, but for the light quarks PDF's we have taken $\protect\mu %
^{2}=-k_{2}^{2}$ (the virtuality of the $W$). This is the set of scales
taken by Stelzer et al in \protect\cite{SSW}.}
\label{ptoptrans_rep_+-q2_gl=1_gr=0willcuts}
\end{figure}

\begin{figure}[tbp]
\epsfysize=10.cm 
\centerline{\epsffile{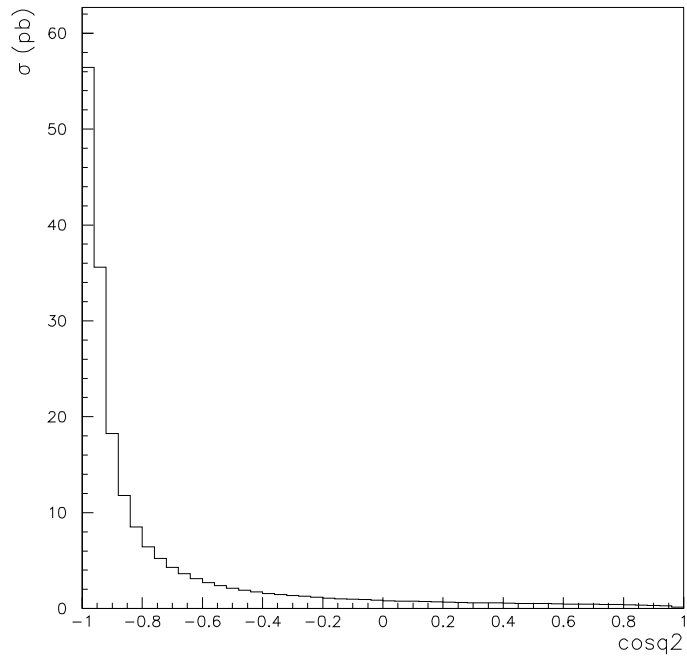}}
\caption{Distribution of the cosine of the angle between the spectator quark
and the gluon corresponding to unpolarized single top production at the LHC.
The momentum of the gluon is in the beam line direction but its sense is not
observable so to obtain an observable distribution we have to symmetrize the
above one. The calculation was performed at the tree level in Standard Model
with $\protect\mu ^{2}=\hat{s}=\left( q_{1}+q_{2}\right) ^{2}$.}
\label{cosq2_lab_unpol_gl=1_gr=0}
\end{figure}

\begin{figure}[tbp]
\epsfysize=10.cm 
\centerline{\epsffile{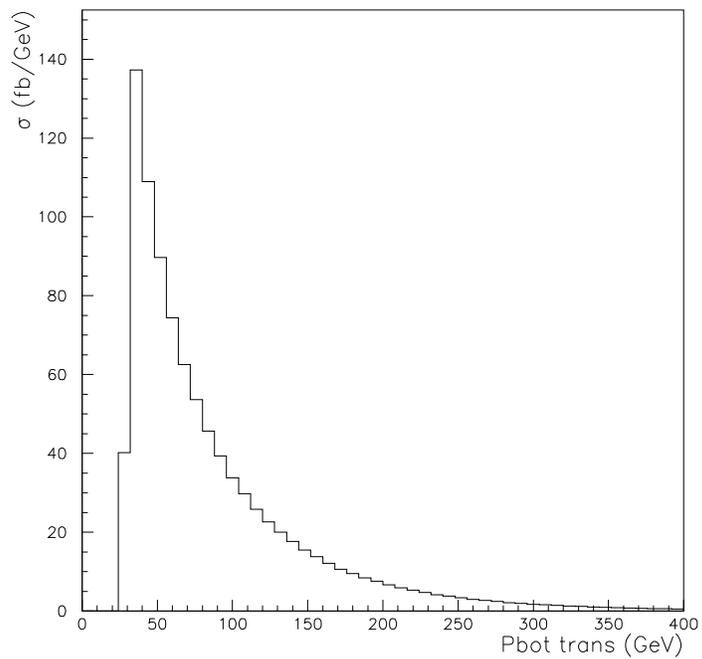}}
\caption{Anti-bottom transversal momentum distribution corresponding to
unpolarized single top production at the LHC. The calculation was performed
at the tree level in the Standard Model. Note the 30 GeV. cut implemented to
avoid large logs due to the massless singularity in the total cross section.
In this plot $\protect\mu ^{2}=\hat{s}=\left( q_{1}+q_{2}\right) ^{2}$ too.}
\label{pbottrans_lab_unpol_gl=1_gr=0}
\end{figure}

\end{document}